\documentclass[aip,reprint]{revtex4-1}

\usepackage{graphicx}
\usepackage{dcolumn}
\usepackage{bm}
\usepackage{mathrsfs}
\usepackage{natbib}
\usepackage{amsmath}
\usepackage{amssymb}
\usepackage{Jonasmacros}
\usepackage{mathptmx}
\usepackage{pifont}
\usepackage{times}
\usepackage{txfonts}

\usepackage[colorlinks,citecolor=magenta,linkcolor=red]{hyperref}
\usepackage[usenames,dvipsnames,svgnames]{xcolor}

\date{\today} 
\begin{document}

\title{Should it really be that hard to model the chirality induced spin selectivity effect?}

\author{Jonas Fransson}
\email{Jonas.Fransson@physics.uu.se}
\affiliation{Department of Physics and Astronomy, Box 516, 75120, Uppsala University, Uppsala, Sweden}

\begin{abstract}
The chirality induced spin selectivity effect remains a challenge to capture with theoretical modeling. While at least a decade was spent on independent electron models, which completely fail to reproduce the experimental results, the lesson to be drawn out of these efforts is that a correct modeling of the effect has to include interactions among the electrons. In the discussion of the phenomenon ones inevitably encounters the Onsager reciprocity and time-reversal symmetry, and questions whether the observations violate these fundamental concepts, or whether we have not been able to identify what it is that make those concepts redundant in this context. The experimental fact is that electrons spin-polarize by one or another reason, when traversing chiral molecules. The set-ups are simple enough to enable effective modeling, however, overcoming the grand failures of the theoretical efforts, thus far, and formulating a theory which is founded on microscopic modeling appears to be a challenge. A discussion of the importance of electron correlations is outlined, pointing to possible spontaneous breaking of time-reversal symmetry and Onsager reciprocity.
\end{abstract}

\maketitle

\section{Introduction}
Who would have thought twenty years ago that the electron spin may actually have something to do with the way life is constructed here on planet earth. I have certainly always thought that the electron spin, and the spin of any quantum particle for that matter, was important only in the context of building up matter to become stable. Typically, electrons pair up in molecules and matter in ways that prohibit chemical reactions, or at least very rapid such – think of explosions. Some reactions do occur spontaneously of course, however, those are mostly either transient or having to do with living organisms which are also transient.

Since electrons and their spin are present literally everywhere in our universe and on our planet earth, the fact that electrons and their spin are also part of the living life is a truism. However, the way I am thinking about it here, is not the banal truth but rather that spin may have played an active role in the formation of molecules that eventually evolved into ribonucleic acid and deoxyribonucleic acid, or, RNA and DNA for short. While nothing can be definitively certain about what took place on the surface of our planet some four to four and a half billion years ago, the best that can be done in a laboratory environment is to establish a viable sequence of chemical reactions that evolves up to either RNA itself or at least some precursor molecule that is known to constitute a foundational step for the origins of life. Quite unexpectedly, this is where a physical phenomenon coined as the chirality induced spin selectivity effect may make a significant difference in the prebiotic chemistry.

Life as we know it is possible due to homochirality \cite{PNAS.2204765119}. Natural aminoacids and proteins are all left handed and the sugers we make use of for our living are all right handed. Homochirality is necessary to enable reproductive and replicative chemical processes. Recall that, for instance, the DNA replicates itself as the body grows and needs more cells.

It has been hypothesized that the chiral RNA precursor molecule ribose-aminooxazoline (RAO) is purified through reciprocal action with magnetite, Fe$_3$O$_4$, which is a ferrimagnetic mineral that has been present on earth from the beginning \cite{SciAdv.9.eadg8274,NatComms.14.6351}. Laboratory measurements showed that RAO molecules selectively adsorb on a ferromagnetic surface in the sense that one enantiomer binds faster and stronger to a surface with a given ferromagnetic orientation compared with the other enantiomer. Furthermore, when RAO is adsorbed to the ferromagnetic surface, the ferromagnetic domain increases as does its coercive field \cite{NatComms.14.6351,arXiv:2412.05720}. The interplay is strong enough to completely discriminate one enantiomer during the course of, say, a hundred million years. There has to be an initial asymmetry of the magnetite magnetization, an asymmetry which is possibly caused by the earth magnetic field.

I suppose that we will never know for sure whether the speculations outlined above has anything to do with the origins of life, but they certainly serve the purpose of bringing reason into the picture. A rational narrative can be constructed and perhaps that is sufficient insofar the story of our origin is concerned. Regardless, having placed the chirality induced spin selectivity effect at the center of the  tale of evolution makes it ever so interesting to develop a theoretical understanding of the phenomenon itself. And should it not, in the end, be a viable effect for the origin of homochirality, there are plenty other potential and realistic applications of the chirality induced spin selectivity effect to motivate understanding it, unless the effect itself is not interesting enough.

How about electrocatalysis \cite{JPhysChemC.123.3024,JPhysChemC.124.22610}, oxygen reduction and evolution reactions, e.g., respiration and water splitting \cite{JPhysChemC.123.3024,JPhysChemC.124.22610,PNAS.30.2202650119,JPhysChemLett.14.1756}, formation of stable molecular spin configurations \cite{NatComms.8.14567,NanoLett.19.5167}, spin-galvanic effect (current induced spin-polarization) \cite{JPhysChemLett.15.6370}, and enhancing thermal and magnetic stability of ferromagnetic metals \cite{NatComms.14.6351,arXiv:2412.05720,JPhysChemLett.16.2001}. Just to mention a few areas where the chirality induced spin selectivity effect has proven its potential to make a difference.

\section{Challenges}
On the one hand, the chirality induced spin selectivity effect \cite{Science.283.814,Science.331.894} has proven itself to be an outstanding challenge for constructing a viable and acceptable theoretical understanding around. There is a spectrum of theoretical foundations the effect apparently, or at least appears to, violate. Although it has been suggested several times, it has not been reproductively established that the chirality induced spin selectivity effect is necessarily associated with spin-polarized currents. Nonetheless, tracing back in the literature for theoretical modeling and calculations, there are numerous examples of which spin-polarized transmission and current is taken for granted, without establishing what makes the connection between a spin-polarized current and the chirality induced spin selectivity effect\cite{JChemPhys.131.014707,EPL.99.17006,JPCM.26.015008,PhysRevB.88.165409,JChemPhys.142.194308,PhysRevE.98.052221,PhysRevB.99.024418,NJP.20.043055,JPhysChemC.123.17043,PhysRevB.85.081404,PhysRevLett.108.218102,PNAS.111.11658,JPhysChemC.117.13730,PhysRevB.93.075407,PhysRevB.93.155436,ChemPhys.477.61,JPhysChemLett.9.5453,JPhysChemLett.9.5753,JChemTheoryComput.16.2914,CommunPhys.3.178,NewJPhys.22.113023,PhysRevB.102.035431,NanoLett.21.6696,JPhysChemLett.12.10262,NanoLett.21.10423}. In fact, there are plenty of confusing and prejudicial statements circulating which originate from the theory of magnetism, statements which may or may not apply in this context, clouding the obvious observable quantities that are to be considered.

It is, for example, unfortunate that the term \emph{spin-polarization} has become established as the quantity that signifies the chirality induced spin selectivity effect. As discussed in the succeeding section, electron spin-polarization can be directly measured using photo-emission spectroscopy combined with Mott scattering \cite{Science.331.894}. On the other hand, since the effect is also related with measurements of charge transport, the meaning of spin-polarization is unclear since the measurements reflect anisotropic responses to an external magnetized reservoir. The meaning of anisotropic response is that the charge current is a function of the magnetization $M$ in one (or both) reservoir(s) interfaced with the chiral molecule, such that $J=J(M)$. Then, what is often referred to as the spin-polarization is the ratio $[J(M>0)-J(M<0)]/[J(M>0)+J(M<0)]$, which is closer to a magneto-resistance, or -current, than a spin-polarization.

Another issue, one that is also frequently debated, is whether the spin-orbit coupling in organic matter can be large enough to sustain a strong spin-polarization/anisotropic response, see for instance Refs. \citenum{PhysRevLett.108.218102,PNAS.111.11658,PhysRevLett.125.263002,NanoLett.21.10423,NanoLett.24.12133}. While spin-orbit coupling is likely to be crucial for coupling the geometrical structure of the molecule to the electronic spin degrees of freedom, it is questionable whether this is where the focus of theoretical developments should lie. Spin-orbit leads to mixing of the spins, thus undoing the \emph{good quantum number} quality of the spin. In this respect it is unclear why a strong spin-orbit coupling would rationalize a strong spin-polarization/anisotropic response, since spin-orbit coupling typically is detrimental for exactly this.

On the other hand, it is well known that non-interacting channels cannot spontaneously give rise to spin-polarized transport\cite{JPAMT.41.405203}. Unless there are external force fields acting on the electron system, that is. Spontaneous formation of spin-polarization and magnetism requires interactions between the electrons. Whether these are formulated in terms of a Hubbard model like framework, see, e.g., Ref. \citenum{Fradkin2013}, or in other ways is less important. The crucial aspect is the existence of either direct or indirect electron-electron interactions. It is an enigma that it took more than a decade of theoretical efforts before interactions became part of the discussion. In the current debates of viable mechanisms, there are mentions of polarizability and spin-dependent tunneling barriers \cite{JPhysChemLett.9.5453,JPhysChemLett.11.1550,ACSNano.16.18601,NatComms.14.5163}, thermal decoherence and disorder \cite{JPhysChemLett.9.5753}, changes in the work function \cite{ACSNano.18.6028,NatComms.16.37}, violation of Onsager’s reciprocity theorem \cite{NanoLett.19.5253}, and how time-reversal symmetry cannot be broken by the molecular structure \cite{PhysRevB.93.075407,PhysRevB.102.035445}.

One can only agree that there are challenges which are not easily overcome by addressing the problem in a straight forward conventional manner. Something novel has to be inferred in order for the chirality induced spin selectivity effect to be understood in a decent way. And as we, as a community, try to find reason with the physics, we have to be inventive, we probably have to go against conventions and axiomatic statements. Not so much for the sake that these are necessarily incorrect or that they do not apply, more for the sake of finding the missing reasons for why these are perhaps not valid in the context. Or perhaps, prejudicial thinking is what lies between the current theoretical status and true theoretical progress. Could it be that we attribute conditions to a system that are not actually satisfied? Or is it because we have been taught to address problems in certain way it becomes to big of a sacrifice to try other routes? When it was introduced, the mean field theory behind the Bardeen-Cooper-Schrieffer model\cite{PhysRev.106.162,PhysRev.108.1175} was unconventional, yet it has become standard ever since.

At several occasions I have heard colleagues claiming that the chirality induced spin selectivity effect is impossible. That there must be something in the way the observations are interpreted that is wrong. I understand these exclamations as a response to that the results do not match our theoretical understanding of physics. Hearing claims of such type is certainly distressing. And while these claims, or perhaps questionings, are necessary for making progress, they are also quite non-constructive and are a little reminiscent of urge to follow the map when the map and environment do not match. It also sounds like we as a community have not learnt anything from the drastic reformation of physics in the late nineteenth and early twentieth centuries. The comparison is not fair and is not meant to trivialize the advent of quantum mechanics, by saying that the chirality induced spin selectivity will call for a new revolution á la the introduction of quantum mechanics. It will certainly not. One can, nonetheless, be a bit more humble and open minded in front of the unknown and ask for less dogmatism and fundamentalism.

The ways we have accepted to manifest that we understand an issue is first and foremost by simplifying and unifying. Whenever a phenomenon can be sorted into a specific category, this act is accompanied by the perception that we do understand the phenomenon. However, this ontological approach or view of the world has nothing to do with understanding. It is nothing more than categorizing, classifying, the world. By knowing that a piece of matter is ferromagnetic does not tell me anything about the origin of its ferromagnetism. Moreover, even a skilled physicist will not be able to say anything about the origin of its ferromagnetism without being provided additional information. For instance, magnetite was discovered long ago and for long only its macroscopic magnetic properties were known, nothing about the origin of its properties. And even now, when we do have a way to describe the mechanisms that give rise to the ferromagnetic property, we can still ask how deeply we do understand. Uttermost, there are postulates that hold up quantum mechanics. In a sense, what we claim to be an exact description of the natural world ultimately depends on metaphysics.

What I am trying to communicate is that we shall absolutely discuss and debate the ways to best approach and comprehend the physical phenomena that surround us. However, perhaps we should try act less priest like, with less aggressions toward each other and be more humble to the fact that thinking is not a simple trade. Thinking is hard work and requires sacrifices of all kinds from the one that delves deeply into such engagement. One has to remember that thinking has nothing to do with ontological arrangements, however, it may be a way to deceive oneself and others about that something is actually being done. A table or a graph is a tangible product that can be shown and communicated. By contrast, thinking is about re-education, reformation of one's mind, thoughts, and views, which is the very reason as to why thinking is not simple. It challenges everything about oneself, perhaps one has to betray one's convictions and beliefs.

\section{Experiments and theory}

\begin{figure}[t]
\begin{center}
\includegraphics[width=\columnwidth]{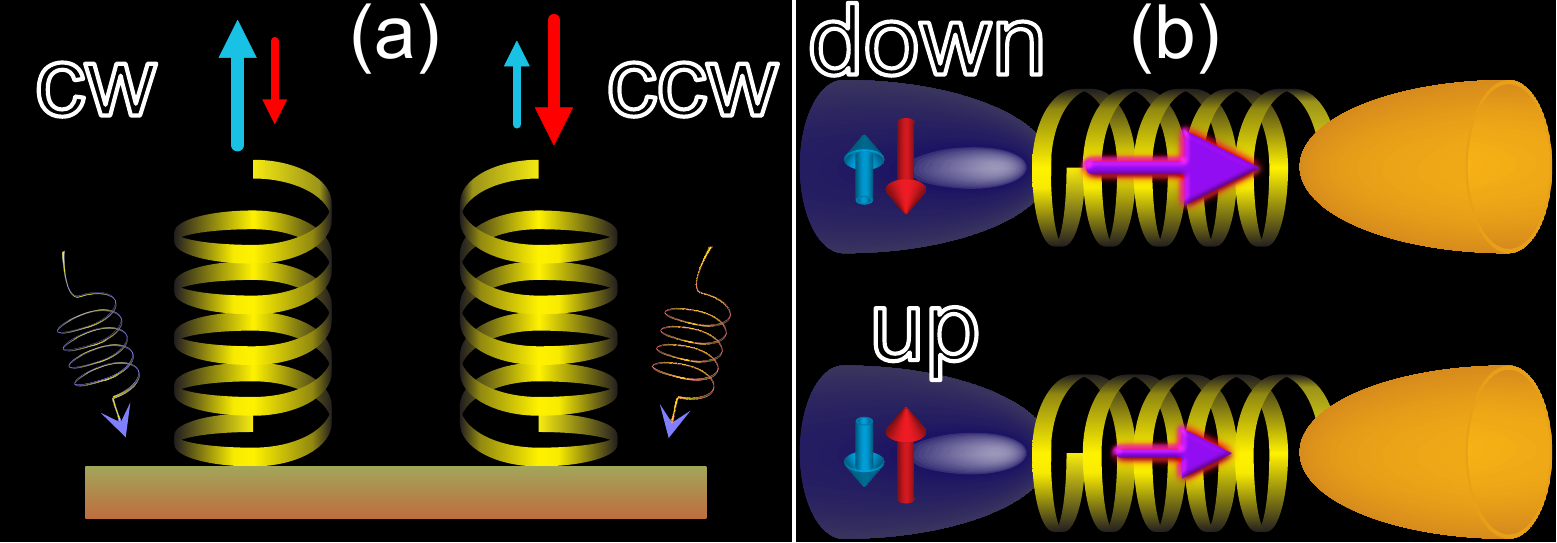}
\end{center}
\caption{{\bf Schematics of the typical experimental set-ups.}
(a) In the photo-emission spectroscopy (PES), the molecule is adsorbed on a metallic substrate which is illuminated with linearly or circularly polarized light. The accelerated electrons are emitted from the loose end of the molecule and their spin-polarization is subsequently detected through Mott scattering. This procedure gives a direct measurement of the spin-polarization of the emitted electrons, hence, the spin selectivity generated in the molecule.
(b) The transport experiments are conducted by measuring the charge current flowing through the molecule mounted between metallic leads, one of which is ferromagnetic. The chirality induced spin selectivity effect is measured indirectly as the response of the charge current to the magnetization of the ferromagnet. Here, the CISS magneto-resistance is used as the measure of spin selectivity.}
\label{fig-schematic}
\end{figure}

Nonetheless, the experimental observations are clear, both the ones resulting from photo-emission spectroscopy (PES) \cite{Science.283.814,Science.331.894,ACSNano.16.12145} and from transport measurements \cite{NanoLett.11.4652,AdvMater.28.1957,Small.20.2308233}. Schematic illustrations of the two set-ups are shows in Fig. \ref{fig-schematic}. While the photo-emission spectroscopy, Fig. \ref{fig-schematic} (a), provides a direct measure of the spin-polarization with which the electrons are emitted from the molecule, the transport set-up, Fig. \ref{fig-schematic} (b), only gives an indirect account of whatever happens in the molecular junction.

Photo-emission spectroscopy in this context is conducted on the composite system where (chiral) molecules are adsorbed on a metallic substrate. By illuminating the substrate, electrons are accelerated in to the conduction, or empty, orbitals of the molecule from which they are eventually emitted into vacuum. The spin-polarization of the emitted electrons is recorded through Mott scattering. Interestingly, in this set-up there is no need for ferromagnets or magnetic fields, not even circularly polarized light, which otherwise is necessary for a spin-dependent light-matter interaction.

While the flux of electrons measured in the photo-emission experiments is spin-polarized and, thereby, constitutes direct evidence of that spin-polarization inevitable must be associated with the chirality induced spin selectivity effect, the challenges with interpreting the outcome in the transport set-up are more intricate. From the results with photo-emission one can, of course, simply infer that the charge current also must be spin-polarized which, therefore, leads to the anisotropic response recorded in the experiments. This is, on the other hand, quite problematic since it offers a metaphysical justification of a spin-polarized transport which may or may not have anything to do with the anisotropic response. Without establishing on theoretical (and mathematical) grounds as to how a spin-polarized current necessarily leads to a anisotropic response in the context of chiral molecules, studies of spin-polarized transmission coefficients can never become more than an academic exercise.

Because of the poor match between spin-polarization and the anisotropic response observed in measurements of the charge current, the term \emph{CISS magneto-resistance}, where CISS is short for chirality induced spin selectivity, has been proposed as a replacement. Therefore, this term is going to be used henceforth in this text.

In what follows, I shall discuss some aspects regarding the chirality induced spin selectivity effect which I have found interesting and compelling. I begin with the Onsager reciprocity and time-reversal symmetry.

\begin{figure}[t]
\begin{center}
\includegraphics[width=\columnwidth]{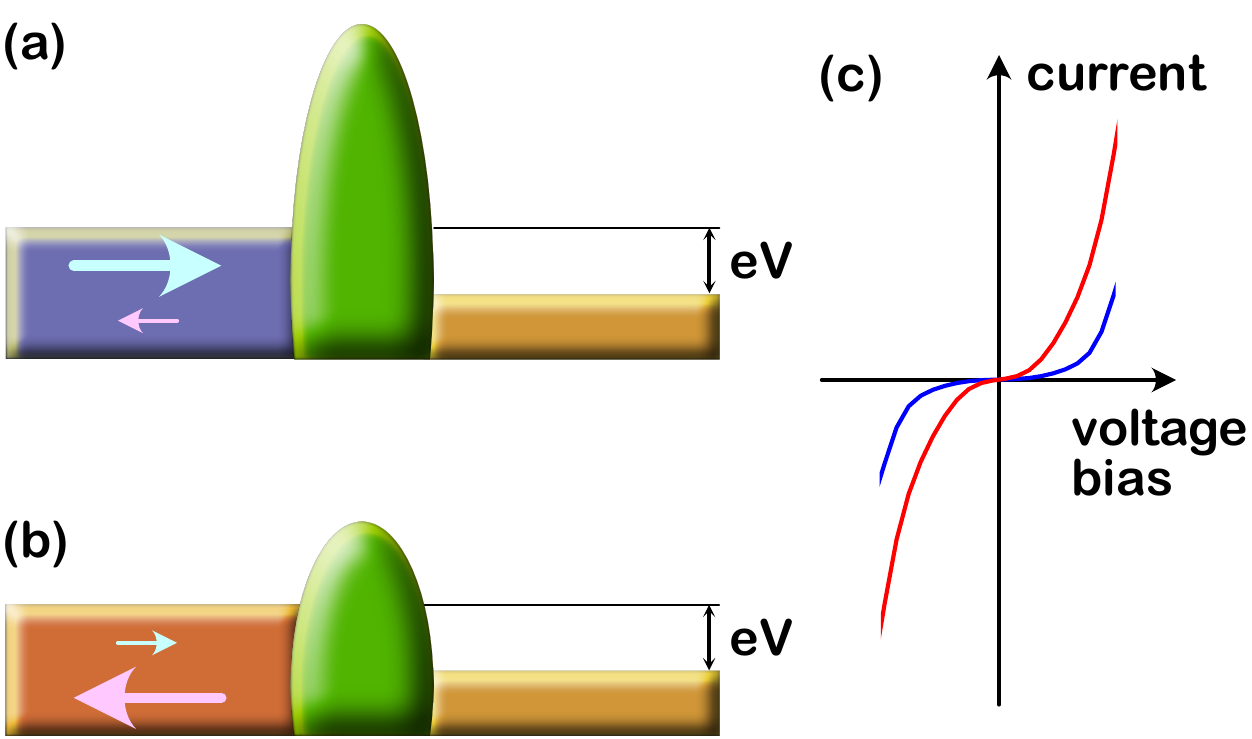}
\end{center}
\caption{{\bf Reductionists view of the CISS magneto-resistance.}
(a), (b) Effective models of the transport set-ups with a ferromagnetic left reservoir and non-magnetic right. The potential barrier between the reservoirs are apparently of different heights, consequently leading to different currents (c) as functions of the voltage bias.}
\label{fig-PotentialBarriers}
\end{figure}

The Onsager reciprocity \cite{PhysRev.37.405} is an interesting issue in the context. In a more general perspective, the Onsager reciprocity theorem tells that thermodynamical quantities, like charge distribution, electrical and thermal conductivities, and Peltier and Seebeck coefficients, in the linear response regime do not vary with the applied field. The applied field may be electrical, magnetic, or thermal, for example, and implies simple linear relationships like Ohm's and Fourier's laws for electrical respectively thermal transport. For the purposes here, the essential knowledge provided by the Onsager reciprocity is that as long as the system remains in the linear response regime, a magnetic field applied, say, parallel or anti-parallel to the voltage bias, does not change its electric resistance. The system pertaining to this scenario is a chiral molecule mounted between two metals, one of the metals sustaining spin-polarized electrons, that is, a ferromagnet. The question is if the Onsager reciprocity theorem is valid in this context? I do not have the answer to this question except that the model I have been utilizing \cite{PhysRevB.102.235416} suggests that the assumptions pertaining to the theorem are not fulfilled. That does not mean anything more than that. It is interesting to notice, however, that others have started out from one position \cite{PhysRevB.99.024418,NatMaterials.5.638,ACSNano.16.18601} and reconciled into another \cite{ACSNano.18.6028,NatComms.14.5163,NatComms.16.37} and suggesting that the potential barrier between the metals changes as function of the magnetization of the ferromagnetic metal, see Fig. \ref{fig-PotentialBarriers}. The figure illustrates a reductionists view of the effective transport properties, or potential barriers, resulting from, panels (a), (b), different spin-polarizations in the ferromagnetic reservoir, corresponding to the measured currents, panel (c). These suggestions are surely interesting and reasonable, however, since they are invoked without reference to any type of microscopic description, more work along those directions is necessary to evaluate the sense in them.

Time-reversal symmetry presents a far more problematic discussion. Is there anything in the molecule--metal heterostruture that can break it? Again, keeping in mind the chiral molecule mounted on one or between two non-magnetic metals. Most, if not all, physicists would say that there is nothing in this set-up that can break time-reversal symmetry. Even some colleagues will burst out that the effect recorded in the experiments, such as the anomalous Hall effect\cite{NatComms.8.14567,JPhysChemLett.10.1139} and induced magnetization\cite{JPhysChemLett.10.3858,Small.20.2406631} are impossible because of time-reversal symmetry and the Onsager reciprocity. Obviously, something is wrong since we cannot agree even on whether the experimental results are physically feasible and, given that they, nevertheless, are reproductively recorded, how to interpret the physics revealed in the measurements.

The transport set-up has a few challenges that have to be addressed separately from the discussion of a spin-polarization. The symmetry breaking agent, for instance, is a magnetized lead which magnetization can be switched by external control, see Fig. \ref{fig-schematic} (b). It is not obvious, however, that the charge current should vary as function of the magnetization since it would require that changes in the transport properties are induced by switching the magnetization. The transport properties are essentially captured in the current-current correlation function, in terms of which the Kubo formula for electrical conductivity is expressed\cite{JPhysSocJpn.12.570}. It should be stressed, however, that the Kubo formula is a linear response result, such that the Onsager reciprocity theorem applies. Recall that a linear response theory implies that the properties of the system of interest do not change under the influence of external force fields. Whether the fields are of electric, magnetic, thermal, or mechanical nature should, in principle, not matter. Consequently, in linear response, the electrical conductivity cannot change as function of the magnetization of the reservoir. However, since the experimental observations clearly suggest that exactly this occurs, the Kubo formula cannot be expected to be useful in the present context.

Onsager reciprocity requires, moreover, that time-reversal symmetry is maintained. In principle one may argue that since the system comprises a magnetized reservoir this requirement is not fulfilled and that there is no problem. However, that is to oversimplify the conditions of the set-up since Onsager has a response even to this issue. By switching the magnetization and simultaneously switching the direction of the electron's motion the amplitude of the charge current should, in the linear respons regime, remain intact. Again, experiments show that this is not the case.

The major problem faced when constructing models for the chirality induced spin selectivity effect is, therefore, that experimental observations suggest the absence of a linear response regime with respect to the magnetized reservoir.

\section{Discussion of a few concepts}
Unfortunately, the state of the theoretical progression was for a long time concerned with the issue of a spin-polarized transmission without even attempting a connection to the CISS magneto-resistance. Thankfully, this attitude towards to the phenomenon of the chirality induced spin selectivity effect has lately been much abandoned to the benefit of targeting the CISS magneto-resistance. In doing so, it follows that nothing can really be achieved without inclusion of effective electron-electron interactions \cite{NanoLett.19.5253,JPhysChemLett.10.7126,PhysRevB.102.214303,PhysRevB.102.235416,JACS.143.14235,JPhysChemC.125.23364,JPhysChemC.127.6900,JPhysChemLett.14.340,JPhysChemC.128.6438,NanoLett.24.12133}. And this conclusion is not taken out of thin air since the molecular state is never an independent particle state but a truly correlated many-particle state. Thus, the reductionist view that the effect of chirality can be compressed into the transmission without gathering a greater picture and understanding what the transport measurements actually encompass, will inevitably fail. By overlooking simple and fundamental facts about the molecular state, the starting point is going to be erroneous from which any type of result can be derived.

The perception that the two enantiomers of a chiral molecule respond anisotropically to the magnetization of a ferromagnetic reservoir has been accommodated in Refs. \citenum{NatComms.14.5163,ACSNano.18.6028,JPhysChemLett.10.7126,PhysRevB.102.235416,JPhysChemLett.16.4346}. In the first two of these references, the effective tunneling barrier between the metals is either assumed to be spin-dependent \cite{NatComms.14.5163}, which consequently is different for the two enantiomers, or arise from different work functions \cite{ACSNano.18.6028} for the two structures. In both suggestions, the magneto-resistance is a repercussion of enantiospecific details of the tunneling barrier, however, in none of these there is any discussion of the mechanisms that lead to this enantiospecificity.

Despite the lack of microscopical modeling of the assumed conditions, these proposals do point to the fact that something in the set-up has to depend strongly on the boundary conditions, that is, the magnetization of the ferromagnetic reservoir, and the chirality of the molecule in the junction. Viewed in this respect, both proposals likely correctly capture the physics. The lack of microscopical detail is, nonetheless, disturbing.

An attempt along the direction of constructing a model description of the physics in the molecular junction is, however, provided in Refs. \citenum{JPhysChemLett.10.7126,PhysRevB.102.235416,JPhysChemLett.16.4346}, and further developed in Refs.  \citenum{JPhysChemC.125.23364,JPhysChemC.127.6900,JPhysChemLett.14.340,NanoLett.24.12133,JPhysChemLett.16.9107}. The model builds on the concept that electron correlations, provided through the coupling between the electrons and nuclear vibrations, introduces this anisotropic response to the external magnetization. The question is what it is in this model that generates an appreciable magneto-resistance. It turns out that including only either interactions or reservoirs which can absorb leakages, is not enough to generate the magneto-resistance. Even so, including both interactions and reservoirs is not sufficient, as well. It is necessary that the two mechanism are joined together, and not only indirectly as a coupled system, but the leakage to the environment, whether it being a thermal reservoir or the leads, makes a crucial difference on the impact of the interactions. It is shown in Ref. \citenum{JPhysChemLett.16.4346} that a coupling of the interactions between the electrons and nuclear vibrations to a thermal reservoir breaks time-reversal symmetry in a fashion that connects to the spins of the electrons in the set-up. It is also shown that replacing the nuclear vibrations with an on-site Coulomb repulsion such that the effect of the repulsion impacts the level broadening caused by the reservoirs, leads to the same type of physics -- underscoring that it is the spin-exchange associated with the effective electron-electron interactions that plays an important role for capturing the physics. That is, whether it is direct Coulomb repulsion or vibrationally assisted indirect attraction between the electrons does not matter for causing the chirality induced spin-selectivity effect. The key aspect is that interactions are included. Moreover, the interactions have to couple directly with the states in the reservoirs which facilitates the anisotropic response to the magnetization of the ferromagnet.

Because, it should kept in mind that it is not only by changing the enantiomers that show difference in the transport properties. Likewise, by switching the magnetization of the ferromagnetic metal, the charge current flowing across the molecular junction changes significantly, meaning everything between one or a few percents to nearly a hundred. Whatever the cause of the magneto-resistance, there are not many types of mechanisms that can be included or excluded in the modeling which leads to that the molecule responds asymmetrically to the magnetization of the ferromagnet. The two set-ups, in which the ferromagnet is magnetized in one or the other direction, represent two essentially different systems, for which the same differential equation can be formulated, however, with two distinct boundary conditions. The boundary conditions separates the two systems, however, they do not make any difference should the equation be linear. With a non-linear equation, the boundary condition can make such a difference that it projects the solutions into markedly different regimes of the phase space. Whether this is relevant in the context of the chirality induced spin selectivity effect is not clear but cannot be dismissed without investigation.

Then, coming back to the time-reversal symmetry breaking. Perhaps it is necessary that time-reversal symmetry is broken for the chirality induced spin selectivity effect to make sense. Perhaps it not necessary. The question still remains: what allows for the dramatic changes in the resistivity when the magnetization of the ferromagnet is switched? The observations of spin-polarized electrons in the photo-emission measurements is a direct indication that there is an imbalance between the spin-projections of the ejected electrons. Such imbalance is conventionally perceived as a manifestation of broken time-reversal symmetry. Different charge currents as a response to changed spin-polarization in the reservoir may, on the other hand, be obtained without breaking the same symmetry. Since the amplitude of the charge current depends on the distribution of occupied and unoccupied states in the system, there could be various reasons for these distributions to respond unequally to the external spin-polarization without themselves becoming spin-polarized. Not that I know of any good such reason, but that does not mean they do not exist. Attributing the chirality induced spin selectivity effect to the electron spin is, nonetheless, the most straight forward and natural accessory despite the potential conflict with a great portion of general physical knowledge.

Whether the issue of time-reversal symmetry can be understood by introducing interactions or not, there is a fundamental reason for why electron-electron interactions are important in the discussion. The chiral molecules considered in the context of the chirality induced spin selectivity effect are closed shell systems. Hence, the molecular states accessed via the transfer of electrons through the molecule are singlet states. It is not by accident or sloppiness this vernacular has been introduced, it has the true connotation of a singlet state. As such, it is a correlated state, with the meaning that the electrons constituting this state are dependent on each other. Independent electrons will always form a Kramers doublet whereas correlated electrons form singlets, triplets, and higher order multiplets. In the adsorption process, mounting the molecule onto one or two metals, the system cannot any longer be considered as a molecule and metals separated from each other. The molecule and the metals are now part of a composite system. More importantly, the molecular component has to be considered as an open shell structure where electrons can leak between the molecule and the metals.

\section{Aspects of modeling}
For the sake of being more concrete regarding how the modeling may be conducted in this context, a fairly general view is introduced and discussed here. It should be remarked already from the beginning that this proposal, like any other, has its pros and cons, however, comprises a necessary set of properties that make the difference between the ability to explain the underlying physics and the ability to literally say nothing at all.

The basic set-up can be found in, e.g., Ref. \citenum{PhysRevB.102.235416}, and is repeated here for convenience. The molecule consists of $\mathbb{M}$ sites embedded in three dimensional space. In a simple helix, for example, the sites are distributed along the coordinates $\bfr_m=(a\cos\varphi_m,a\sin\varphi_m,c_m)$, $1\leq m\leq\mathbb{M}$, where $a$ is the radius, $\varphi_m=2\pi(m-1)/(\mathbb{M}-1)$, and $c_m=c(m-1)/(\mathbb{M}-1)$.

Making a simple set-up, there is an energy level $\dote{m}$ associated with each site $m$, nearest neighbor and next-nearest neighbor mixing with rates $t_0$ and $\lambda_0$, respectively. The latter is, moreover, connected with a spin-orbit kind of coupling of the form $i\lambda_0\bfv_m^{(s)}\cdot\bfsigma$, $s=\pm1$, where $\bfv_m^{(s)}$ accounts for the curvature defined between the sites $m$, $m+s$, and $m+2s$, and $\bfsigma$ is the vector of Pauli matrices. An explicit form of this curvature is given by the vector product $\bfv_m^{(s)}=\hat\bfd_{m,s}\times\hat\bfd_{m+s,s}$, where $\bfd_{m,s}=\bfr_m-\bfr_{m+s}$ is the distance vector between the sites $m$ and $m+s$, and $\hat\bfd_{m,s}=\bfd_{m,s}/|\bfd_{m,s}|$.

Letting the spinor $\Psi=(\psi_1,\psi_2,\ldots,\psi_\mathbb{M})^t$ with $\psi_m=(\psi_{m\up}\ \psi_{m\down})^t$, represent the annihilation of an electron within the molecule, the single-electron Hamiltonian $\Hamil_0$ corresponding to the discussion can be written $\Hamil_0=\Psi^\dagger H_0\Psi$, where
\begin{align}
H_0=&
	\sum_{m,s=\pm1}
	\Bigl(
		-t_0\delta_{mm+s}
		+
		i\lambda_0\bfv_m^{(s)}\cdot\bfsigma\delta_{mm+2s}
	\Bigr)
	.
\end{align}
Electron correlations are, furthermore, introduced via the Hamiltonian $\Hamil_1=\Psi^\dagger\sum_\nu H_\nu Q_\nu\Psi+\Hamil_\text{vib}$, where $H_\nu$ is obtained from $H_0$ by replacing the subscript 0 with the vibrational mode number $\nu$, whereas $Q_\nu=b_\nu+b^\dagger_\nu$ denotes the nuclear displacement of the mode $\nu$ and $\Hamil_\text{vib}=\sum_\nu\omega_\nu b^\dagger_\nu b_\nu+\sum_{\nu\nu'\nu''}\Phi_{\nu\nu'\nu''}Q_\nu Q_{\nu'}Q_{\nu''}$ defines the total unperturbed vibrational energy. The second term in the model for the nuclear vibrations accounts for anharmonic corrections to the harmonic oscillators, which essentially should be included for any structure with broken inversion symmetry, like a chiral. Hence, the model for the molecule is given as the Hamiltonian $\Hamil_\text{mol}=\Psi^\dagger(E+H_0+\sum_\nu H_\nu Q_\nu)\Psi+\Hamil_\text{vib}$, where $E=\diag{\dote{1},\dote{2},\ldots,\dote{\mathbb{M}}}{}$ is the diagonal matrix of the unperturbed on-site energies.

The molecule is, furthermore, interfaced with one or several electron reservoirs $\Hamil_\chi=\sum_{\bfk\in\chi}\dote{\bfk}\psi^\dagger_\bfk\psi_\bfk$, where $\psi_\bfk$ is the spinor for an electron at the energy $\dote{\bfk}$ and crystal momentum $\bfk$ in the reservoir $\chi$. The interface between the molecule and, e.g., two reservoirs, referred to as the left ($L$) and right ($R$), is captured in $\Hamil_T=\sum_{\bfk\in L}\psi^\dagger_\bfk\bfu_{\bfk1}\psi_1+\sum_{\bfk\in R}\psi^\dagger_\bfk\bfu_{\bfk\mathbb{M}}\psi_\mathbb{M}+H.c.$.

This model is simplistic in the sense that it does not capture any realistic molecule \emph{per se}, however, it contains the salient features that are necessary for capturing the relevant physics of the set-up. In particular, the electron correlations are here addressed in terms of a coupling between the electronic and nuclear vibrational degrees of freedom. In order to see that these play an analogous role as a direction Coulomb interaction, it is instructive to consider the following.

The Hamiltonian $\Hamil_1$ can be rewritten as
\begin{align}
\Hamil_1=&
	\sum_\nu\omega_\nu B^\dagger_\nu B_\nu
	-
	\sum_\nu
		\frac{1}{\omega_\nu}
		\Bigl(
			\Psi^\dagger H_\nu\Psi_\nu
		\Bigr)^2
	,
\label{eq-effUmodel}
\end{align}
where $B_\nu=b_\nu+\Psi^\dagger H_\nu\Psi$ is a shifted operator which may be integrated out from the model. Thereby, the physics is transformed into a purely Fermionic model with effective electron-electron interactions -- second term in Eq. \eqref{eq-effUmodel}. Since both repulsive and attractive interactions lead to correlated states in the molecular structure, the origin of the correlations is immaterial for the discussion here.

However, capturing the molecular electron correlations to one or another degree is not sufficient for capturing the chirality induced spin selectivity effect. Namely, the effect of whatever processes that may occur in the molecule when it is isolated from the environment, will be cancelled by compensating processes such that any required conservation law is maintained. Simply put, as long as the molecule is not interfaced with an environment, there cannot emerge any phenomenon that requires spontaneous symmetry breaking. This can only happen if the molecule is interfaced with an environment with macroscopic degrees of freedom. As a comparison, consider the ordered state of a spin-density wave which is described very well by the Hubbard model for a correlated crystal\cite{Fradkin2013}. Thanks to the macroscopic measure of the crystal, an ordered spin density wave state can be stabilized which is also more natural than the paramagnetic state. In the case of a molecule, the analogous macroscopic measure can be provided by the reservoir.

Even without alluding to the results related to the Hubbard model, it is not terribly involved to demonstrate that a model of the type $\Hamil=\Hamil_L+\Hamil_R+\Hamil_T+\Hamil_\text{mol}$ leads to a state of a chiral molecule which has a resemblance with the spin density wave, a result which can be understood as a locking of the fluctuating spin singlet configurations into one specific which is determined, or protected, by chirality. One may think of the system as to reach a stationary state where all spin configurations but one have been damped out by the presence of the reservoirs.

Without going into the deeper details of the calculations, which are presented in \cite{arXiv:2509.17817}, one may show by perturbation theory that charge and spin densities at site 1, which is adjacent to the left reservoir, can be written as
\begin{subequations}
\label{eq-ns1}
\begin{align}
\av{n_1}=&
	\av{n_1}^{(0)}
	-
	\frac{128}{\Gamma^L}
	\frac{\bfp_L\cdot\bfv_1^{(+)}}{(1-p_L^2)^2}
	\sum_\nu
		\calT_\nu
		\coth\frac{\beta\tilde\omega_\nu}{2}
	,
\label{eq-n1}
\\
 \av{\bfs_1}\approx&
 	\av{\bfs_1}^{(0)}
	+
	\frac{32}{\Gamma^L}
	\biggl(
		\frac{\bfv_1^{(+)}}{1-p_L^2}
		+
		\frac{\bfp_L\cdot\bfv_1^{(+)}}{(1-p_L^2)^2}
		\bfp_L
	\biggr)
	\sum_\nu
		\calT_\nu
		\coth\frac{\beta\tilde\omega_\nu}{2}
	,
\label{eq-s1}
\end{align}
\end{subequations}
where $\av{n_1}^{(0)}$ and $\av{\bfs_1}^{(0)}$ are the corresponding unperturbed densities, whereas $\calT_\nu=t_\nu^2/\lambda_0|\bfv_1^{(+)}|^2-t_\nu\lambda_\nu/t_0$, $\tilde\omega_\nu$ is the effective vibrational energy in the anharmonic oscillator, and $1/\beta=k_BT$ is the thermal energy in terms of the Boltzmann constant $k_B$ and temperature $T$. An importance quantity in this expression is the coupling $\Gamma^L$ between the molecule and the left reservoir and which has the functional form $\bfGamma^L=\Gamma^L(\sigma^0+\bfp_L\cdot\bfsigma)/2$, where $\sigma^0$ is the two-dimensional identity matrix and $\bfp_L$ gives a parametrized form of the spin-polarization in the reservoir, such that $|\bfp_L|\leq1$. Formally, this coupling is defined by $\bfGamma^L=\sum_{\bfk\in L}\bfu_{\bfk1}^\dagger(-2\im\bfg_\bfk^r)\bfu_{\bfk1}$, where $\bfg_k^r$ is the retarded Green function for the electrons in the reservoir.

Two important aspects of the results in Eq. \eqref{eq-ns1} has a direct impact on the capability of the model to reproduce the signatures of the chirality induced spin selectivity effect, namely (\emph{i}) there is a non-trivial spin-density $\av{s_1}\neq0$ for all $\bfp_L$, including a non-magnetic reservoir for which $\bfp_L=0$, and (\emph{ii}) the charge density depends linearly on the spin-polarization $\bfp_L$ in the reservoir. The first statement implies that the molecule, when interfaced with the reservoir, assumes a non-trivial spin configuration. Hence, the system spontaneously breaks time-reversal symmetry. The second statement means that the system continuously changes its charge state as a linear function of the spin-polarization in the reservoir. This dependence equivalently means that a linear response regime cannot be defined with respect to the spin-polarization. Uttermost the two statements imply that the basic assumption of microscopic reversibility required for the Onsager reciprocity to apply is not fulfilled. Moreover, since the charge distribution varies with the spin-polarization of the reservoir, an accompanied variation of the charge current can be expected.

The presence of the reservoir, signified through $\bfGamma^\chi$, is not innocent. On the contrary, the result deeply depends on that the molecule is coupled to the reservoir. This coupling puts the closed and finite molecular system into a macroscopic context such that the degrees of freedom of the reservoir becomes accessible to the electronic properties of the molecule. Thanks to the reservoir, all spin-configurations but one are damped out, simultaneously stabilizing the one which is favored by chirality. The effect of this damping mechanism can is quantified by the product between the spin-orbit coupling $i\bfv_m^{(s)}\cdot\bfsigma$ and the level broadening $-i\bfGamma^\chi/2$, giving
\begin{align}
\frac{\Gamma^\chi}{2}
	\bfv_m^{(s)}\cdot\bfsigma
	\Bigl(
		\sigma^0+\bfp_\chi\cdot\bfsigma
	\Bigr)=&
	\frac{\Gamma^\chi}{2}
	\bfv_m^{(s)}\cdot
	\Bigl(
		\bfp_\chi
		+
		\bfsigma
	\Bigr)
	,
\end{align}
where a term of no importance has been omitted. The two terms in this product occur in Eq. \eqref{eq-ns1} as well as in the corresponding expressions for the other sites. As is clear, this product comprises the statements (\emph{i}) and (\emph{ii}) above. In addition, pertaining to the molecular curvature, the second term, $\bfv_m^{(s)}\cdot\bfsigma$, expresses a chirality induced Zeeman splitting and the first term, $\bfv_m^{(s)}\cdot\bfp_\chi$, shows the dependence on the spin-polarization in the reservoir.

\begin{figure*}[t]
\begin{center}
\includegraphics[width=\textwidth]{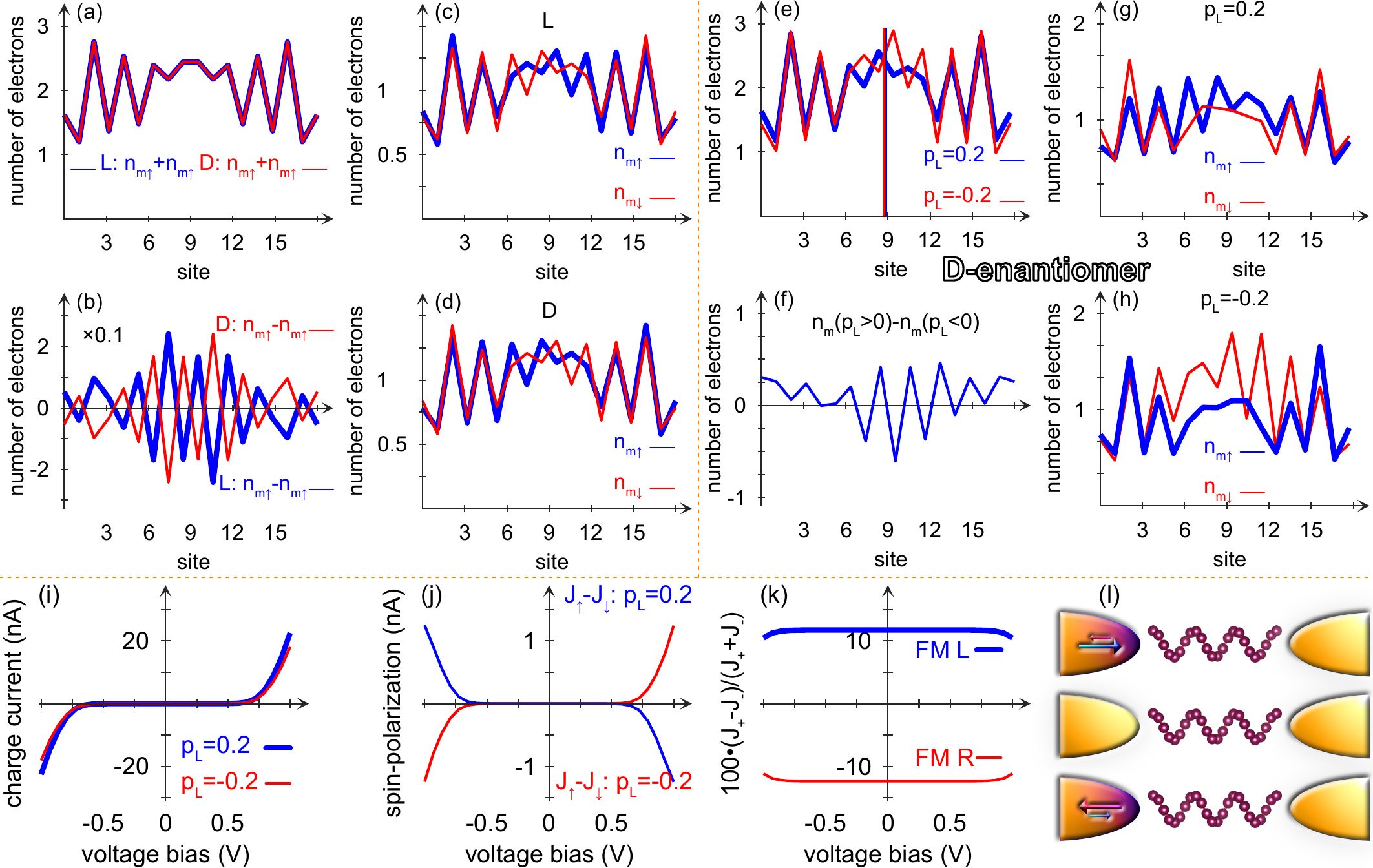}
\end{center}
\caption{
{\bf Calculated electronic properties of chiral molecule reservoir heterostructure.}
Panels (a)--(h) display the site resolved densities for various situations, where (a)--(d) are with non-magnetic and (e)--(h) for ferromagnetic left reservoir. 
(a) Charge densities $n_{m\up}+n_{m\down}$ and (b) corresponding spin densities $n_{m\up}-n_{m\down}$ for (blue) $L$ and (red) $D$ enantiomer. (c), (d) Corresponding spin resolved charge densities for (c) $L$ and (d) $D$ enantiomers showing (blue) $n_{m\up}$ and (red) $n_{m\down}$.
(e) Charge densities for the $D$-enantiomer for a ferromagnet left reservoir with spin-polarization (blue) $\bfp_L=0.2\hat{\bf z}$ and (red) $\bfp_L=-0.2\hat{\bf z}$. The vertical lines indicate the positions of the charge center-of-mass.
(f) Difference between the charge densities $n_m=n_{m\up}+n_{m\down}$ in panel (e).
(g), (h) Spin resolved charge densities for (g) $\bfp_L=0.2\hat{\bf z}$ and (h) $\bfp_=-0.2\hat{\bf z}$.
(i) Charge current for (blue) $\bfp_L=0.2\hat{\bf z}$ and (red) $\bfp_=-0.2\hat{\bf z}$, 
(j) corresponding spin-polarizations $J_\up-J_\down$, and
(k) normalized CISS magneto-resistance for ferromagnetic (blue) left and (red) right reservoir.
(l) Schematics of the chiral molecule mounted in the junction between a (top, bottom) ferromagnetic and (middle) non-magnetic reservoir to the left and non-magnetic to the right.
The calculations are made for helices with $3\times6$ sites, using $\lambda_0=1/50$, $t_\nu=1/20$ and $\lambda_\nu=1/500$ for a single $\nu$, $\Gamma_0=1$, $\dote{m}-\mu_0=-10$, $\omega_\nu=1/50$, and $\Phi=3/500$ in units of $t_0=0.1$ eV, calculated for $T=300$ K.
}
\label{fig-summary}
\end{figure*}

It should be remarked here that the results in Eq. \eqref{eq-ns1} are not singular in the limits $t_0\rightarrow0$ and $\lambda_0\rightarrow0$, since the limits cannot be taken independently of the other parameters in the model. It should also be remarked that the results for the site coupled to the right lead is formally the same as the ones given in Eq. \eqref{eq-ns1}, with the replacements $\bfp_L\rightarrow\bfp_R$ and $\bfv_1^{(+)}\rightarrow\bfv_\mathbb{M}^{(-)}=-\bfv_\mathbb{M}^{(+)}$. The latter replacement indicates that the spin-density at site $\mathbb{M}$ has opposite sign as the spin-density at site 1, thus, indicating a molecular spin-density wave like state, in analogy with the results for the Hubbard model. Further calculations indeed corroborate that the molecular spin structure has the character of a spin-density wave. This can be seen in Fig. \ref{fig-summary} (a)--(d) in which a summary of the calculated charge distributions are shown for both $L$ and $D$ enantiomers embedded in the junction between two non-magnetic reservoirs, corresponding to the middle schematic of the system in Fig. \ref{fig-summary} (l). While the total charge is distributed equally in the two enantiomers, panel (a), the spin projections $n_{m\sigma}$, $\sigma=\up,\down$, manifest broken spin degeneracies which are mirror images of one another in the two configurations, panels (b)--(d). Especially, the non-trivial spin-polarizations plotted in panel (c) appear spin-density wave like with zero net molecular magnetic moment. The parameters used in the calculations of the results in Fig. \ref{fig-summary} are reasonable to realistic molecules \cite{JACS.135.3953,JPhysChemLett.6.4889,JChemPhys.152.214105,JPhysChemLett.9.5753} and are chosen with the ambition to reflect the basic physics rather than for optimized performance. 

The variations of the charge with changed spin-polarization in the left reservoir are well illustrated in Fig. \ref{fig-summary} (e)--(h), showing the charge densities for the $D$-enantiomer $\bfp_L=\pm0.2\hat{\bf z}$, corresponding to the top and bottom schematics in Fig. \ref{fig-summary} (l). The total charge densities, panel (e), are clearly not degenerate for the two spin-polarization, something which is emphasized by the difference between the two conditions plotted in panel (f). In the example, the difference in the local charge is nearly half an electron (site 9). This is further underlined by that the charge polarizations in the two cases are not equal, see the vertical lines in panel (e) which indicate the positions of the charge center-of-mass. The lines are not equal, albeit close to each other. The modified charge densities are accompanied by strongly altered spin densities, panels (g), (h), suggesting that the charge density is closer to being spin degenerate for $\bfp_L=0.2\hat{\bf z}$, panel (g), than it is in the opposite configuration, panel (h). It is, therefore, not a long stretch to predict a larger current in the former case than the latter, since the less it matters which spin the electron has when entering the molecule the larger is the probability for it to contribute positively to the current.

This prediction is corroborated in the transport calculations, see Fig. \ref{fig-summary} (i), showing the charge currents as function of the voltage bias corresponding to the configurations in panel (e). The current for (blue) $\bfp_L=0.2\hat{\bf z}$ has a larger amplitude than for (red) $\bfp_L=-0.2\hat{\bf z}$. The difference between the current becomes clearer by plotting the CISS magneto-resistance, see blue/bold line in Fig. \ref{fig-summary} (k). In agreement with experimental observations, the CISS magneto-resistance is nearly constant over an extended range of voltage biases across equilibrium\cite{NanoLett.11.4652,JPhysChemC.124.10776,NanoscaleHoriz.8.320,Small.20.2308233,JPhysChemLett.15.6605,ACSNano.19.17941}. Interestingly, the modeling also reproduces the recent result from Ref. \citenum{ACSNano.19.17941} showing that the CISS magneto-resistance changes sign when the ferromagnetic reservoir is being switched from the left to the right hand side of the junction, see red line in panel (k). This property can be traced back to the mirror imaged spin configurations in panel (b) -- the $L$ enantiomer has a negative (positive) CISS magneto-resistance for a ferromagnetic left (right) reservoir, thus, fulfilling the complete mirror images of one another.

It may be commented on that the model used for this investigation comprises the coupling between electrons and the nuclear vibrations analogous to the electron-phonon interactions often considered in condensed matter physics. A natural question is, therefore, which relevance chiral phonons have in this context, simply since the molecules are chiral and should nuclear vibrations make a difference it is reasonable to think of these as chiral in one or another sense.

The question can be answered in more than one way. The simplest, but somewhat arrogant, answer is that chiral phonons are not relevant since molecules do not host phonons. Phonons are quantized collective excitations in condensed matter. The closest molecular analoge to the phonons would be coherent molecular vibrations involving most of the molecular structure. For such excitations, it is relevant to think of chiral vibrational modes since such would be collective excitations encompassing the whole chiral structure. Second, for the vibrations to couple to electrons, inversion symmetry typically has to be broken\cite{PhysRevMaterials.1.074404,PhysRevResearch.5.L022039}, a condition which is satisfied in chiral structures. Finally, whether the associated vibrational magnetic moments have any impact on the chirality induced spin selectivity effect remains an open question. It is clear, nonetheless, that chirality impacts the vibrational wave functions as well as the electronic. To which extent the electronic and vibrational degrees of freedom can be decoupled while maintaining a general theory for the chirality induced spin selectivity effect remains unclear.

\section{Applying the chirality induced spin selectivity effect in chemistry}
The theoretical state-of-the-art regarding the chirality induced spin selectivity effect is based on effective modeling and the community has yet to reach a consensus of a viable theoretical description. However, despite the current state of the field, significant theoretical progress is anticipated, as the underlying mechanisms of the chirality induced spin selectivity effect are beginning to be uncovered. For instance, to account for the fact that the molecules are in singlet states, the inclusion of electron correlations and moving away from independent particle theory is necessary. A solid understanding of the fundamental physics underlying the chirality induced spin selectivity effect is essential for elucidating its influence on chemical processes. Emerging theories describing spin-polarization in chiral molecules upon interfacing with metals as well as under non-equilibrium conditions have the potential to enable understanding of the effect of spin on chemical reactions, for example, the oxygen reduction and evolution reactions, occurring on such chiral systems.

While the model Hamiltonians used so far can enable identification of the basic mechanisms underlying the chirality induced spin selectivity effect, first-principles calculations are necessary to obtain quantitative estimates of both the physical and chemical aspects. To this end, present day \emph{ab initio} calculations do not have the capacity to either quantitative or even qualitatively reproduce the physics of the chirality induced spin selectivity effect.

In addition to the electron-electron interactions, another promising pathway forward lies in the incorporation of non-adiabadicity
\cite{PhysRevLett.119.046001,PhysRevB.104.L201409,PhysRevB.106.184302,NatComms.12.700,JPhysChemLett.14.770,JPhysChemLett.14.5665} which has a great potential to add important contributions to the current \emph{ab initio} software. Here, the role of nuclear motion, vibrations, is included by constructing a Berry force which can be incorporated in the electronic single-electron part as a gauge potential $\bfA$, i.e., shifting the momentum operator $p^2$ to $(\bfp-\gamma\bfA)^2$. In this way, the molecular nuclear motion plays a role of an effective magnetic potential which directly modifies the electron wave functions. These developments should possibly advance the field into more quantitative capabilities.

Another feature with chirality that has been raised recently is that the chirality induced spin selectivity effect effect may lead to highly spin-polarized triplet currents \cite{JPhysChemLett.14.1756,JPhysChemLett.14.9377,JPhysChemLett.16.1629,SciAdv.11.eadx4761}, which would also be beneficial in the context of, e.g., catalysis, as this could favor spin aligned intermediates by removing two-electron triplets of a certain spin-projection to the reactant. Extending this model to include the electron transfer from the reactant to the catalyst, both with and without spin polarization of the currents within the catalyst, could allow evaluation of the total energy required for the four electron transfers involved in the oxygen reduction and evolution reactions, and help verify whether this spin polarization can enhance activity. A key advantage of this approach is that it enables the electron transfer kinetics to be examined independently of the other effects.

Moreover, to create understanding of the chirality induced spin selectivity effect on chemical reactions, specific models for the specific reactions should be developed. Since the oxygen reduction and evolution reactions include a process in which the triplet ground state of dioxygen has to be either converted into a singlet or the other way around, the focus in connection with the chirality induced spin selectivity effect has been on these two reactions. This also partly driven by that the questions have been directly related to the supply of electrons for the oxygen reduction reaction to occur, and partly since there are experiments to make direct comparison with, e.g., Refs. \citenum{PNAS.30.2202650119,JPhysChemLett.14.1756,JPhysChemLett.14.9377}.

\begin{figure}[t]
\begin{center}
\includegraphics[width=\columnwidth]{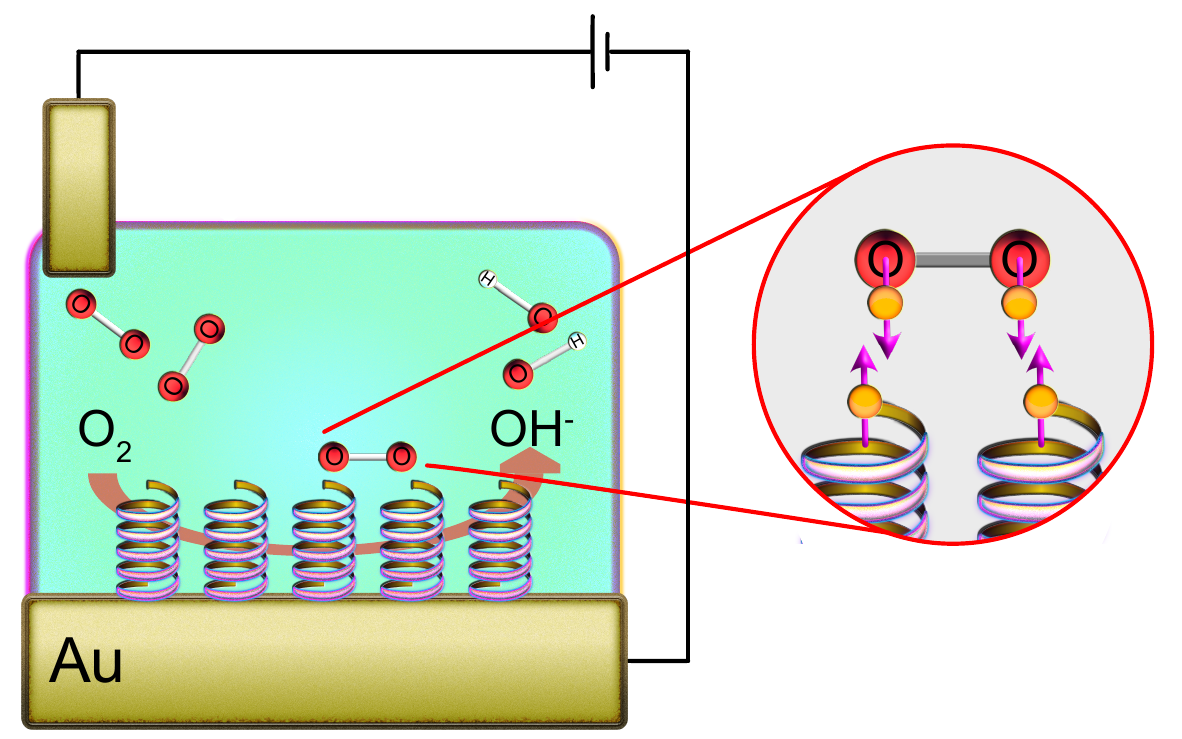}
\end{center}
\caption{
{\bf Schematic of the oxygen reduction reaction catalyzed using chiral molecules.}
By spin selectivity, the chiral molecules provide triplet-like electron pairs with the same spin which neutralizes the angular momentum mismatch between the triplet state O$_2$ and singlet state end products.
}
\label{fig-ORR}
\end{figure}

The reason why chirality induced spin selectivity effect would have anything to do with the oxygen evolution and reduction reactions is that the spin angular momentum mismatch between the reactants is eliminated. The illustration in Fig. \ref{fig-ORR} shows a schematic of the oxygen reduction reaction catalyzed with chiral molecules as demonstrated in Ref. \citenum{PNAS.30.2202650119}. The O$_2$ molecule has a spin triplet ground state whereas typical closed shell structures, inluding H$_2$O, are spin singlets in the ground state. There is, hence, not only a potential energy mismatch but also a mismatch in the angular momentum which has to be overcome by some means. Since the chirality induced spin selectivity according to the photo-emission measurements provides spin-polarized electrons, there is an enhanced probability that pairs of electrons with the same spin are accumulated on the free end of the molecule, see the closeup in Fig. \ref{fig-ORR}. Whether these electrons can be regarded as a true spin triplet or not can certainly be questioned, however, model calculations\cite{JPhysChemLett.16.1629} suggest that the two-electron triplets are strongly spin-polarized and provide enhancements of the two-electron current in chiral molecules compared with achiral. Therefore, a more or less simultaneous transfer of two electrons with the same spin may be sufficient for conserving the total angular momentum when the O$_2$ becomes O$_2^{2-}$ and subsequently dissociates for further reactions.

Although the oxygen reduction and evolution reactions represent opposite directions of the reaction, they involve essentially the same fundamental components -- namely, the redox states of oxygen, catalytic processes, and similar intermediates and ionic species. It can, therefore, be hypothesize that theoretical developments from the study of the oxygen reduction reaction could also guide understanding of the oxygen evolution reaction. Key transferable hypotheses include the idea that simultaneous two-electron transfer could bypass angular momentum constraints \cite{PNAS.30.2202650119,JPhysChemLett.14.1756,JPhysChemLett.14.9377}, while spin-polarized electron removal may reduce activation barriers by circumventing spin selection rules -- potentially enabling access to a triplet reaction surface \cite{PNAS.30.2202650119}. These mechanisms are proposed in the context of the triplet-to-singlet reduction of O$_2$ in both oxygen reduction and evolution reactions, which involves similar spin constraints during the formation of triplet O$_2$. Additionally, defined chirality is hypothesized to lower entropic barriers by limiting the number of accessible reaction states, a concept relevant to both oxygen reduction and evolution reactions \cite{PNAS.30.2202650119}.

\section{Conclusions}
As a concluding remark, in all fairness the decade of theoretical developments in which interactions among the electrons were excluded or omitted, has shown that simplistic non-interacting theories are insufficient for capturing the chirality induced spin selectivity effect. Despite being a negative result, it is an important result. Especially, since the phenomenon that we wish to develop a comprehensive theory for at a simple level appears to break with, what we can call, common sense in physics. It should be a relief to anyone involved that the physics of the chirality induced spin selectivity effect is not that simple that it allows being captured in a simple Landauer-B\"uttiker like transmission of independent channels. For if it was that simple, there would have been troubles elsewhere in the theoretical foundations of physics, problems which would have been far more serious than the ones we are facing in this field. Thanks to interactions, there may be a way for us to find reason with a seemingly unreasonable effect.

\bibliography{CISSref}

\begin{thebibliography}{91}%
\makeatletter
\providecommand \@ifxundefined [1]{%
 \@ifx{#1\undefined}
}%
\providecommand \@ifnum [1]{%
 \ifnum #1\expandafter \@firstoftwo
 \else \expandafter \@secondoftwo
 \fi
}%
\providecommand \@ifx [1]{%
 \ifx #1\expandafter \@firstoftwo
 \else \expandafter \@secondoftwo
 \fi
}%
\providecommand \natexlab [1]{#1}%
\providecommand \enquote  [1]{``#1''}%
\providecommand \bibnamefont  [1]{#1}%
\providecommand \bibfnamefont [1]{#1}%
\providecommand \citenamefont [1]{#1}%
\providecommand \href@noop [0]{\@secondoftwo}%
\providecommand \href [0]{\begingroup \@sanitize@url \@href}%
\providecommand \@href[1]{\@@startlink{#1}\@@href}%
\providecommand \@@href[1]{\endgroup#1\@@endlink}%
\providecommand \@sanitize@url [0]{\catcode `\\12\catcode `\$12\catcode
  `\&12\catcode `\#12\catcode `\^12\catcode `\_12\catcode `\%12\relax}%
\providecommand \@@startlink[1]{}%
\providecommand \@@endlink[0]{}%
\providecommand \url  [0]{\begingroup\@sanitize@url \@url }%
\providecommand \@url [1]{\endgroup\@href {#1}{\urlprefix }}%
\providecommand \urlprefix  [0]{URL }%
\providecommand \Eprint [0]{\href }%
\providecommand \doibase [0]{http://dx.doi.org/}%
\providecommand \selectlanguage [0]{\@gobble}%
\providecommand \bibinfo  [0]{\@secondoftwo}%
\providecommand \bibfield  [0]{\@secondoftwo}%
\providecommand \translation [1]{[#1]}%
\providecommand \BibitemOpen [0]{}%
\providecommand \bibitemStop [0]{}%
\providecommand \bibitemNoStop [0]{.\EOS\space}%
\providecommand \EOS [0]{\spacefactor3000\relax}%
\providecommand \BibitemShut  [1]{\csname bibitem#1\endcsname}%
\let\auto@bib@innerbib\@empty
\bibitem [{\citenamefont {Ozturk}\ and\ \citenamefont
  {Sasselov}(2022)}]{PNAS.2204765119}%
  \BibitemOpen
  \bibfield  {author} {\bibinfo {author} {\bibfnamefont {S.~F.}\ \bibnamefont
  {Ozturk}}\ and\ \bibinfo {author} {\bibfnamefont {D.~D.}\ \bibnamefont
  {Sasselov}},\ }\bibfield  {title} {\enquote {\bibinfo {title} {On the origins
  of life's homochirality: Inducing enantiomeric excess with spin-polarized
  electrons},}\ }\href {\doibase 10.1073/pnas.2204765119} {\bibfield  {journal}
  {\bibinfo  {journal} {Proceedings of the National Academy of Sciences}\
  }\textbf {\bibinfo {volume} {119}},\ \bibinfo {pages} {e2204765119} (\bibinfo
  {year} {2022})}\BibitemShut {NoStop}%
\bibitem [{\citenamefont {Ozturk}\ \emph
  {et~al.}(2023{\natexlab{a}})\citenamefont {Ozturk}, \citenamefont {Liu},
  \citenamefont {Sutherland},\ and\ \citenamefont
  {Sasselov}}]{SciAdv.9.eadg8274}%
  \BibitemOpen
  \bibfield  {author} {\bibinfo {author} {\bibfnamefont {S.~F.}\ \bibnamefont
  {Ozturk}}, \bibinfo {author} {\bibfnamefont {Z.}~\bibnamefont {Liu}},
  \bibinfo {author} {\bibfnamefont {J.~D.}\ \bibnamefont {Sutherland}}, \ and\
  \bibinfo {author} {\bibfnamefont {D.~D.}\ \bibnamefont {Sasselov}},\
  }\bibfield  {title} {\enquote {\bibinfo {title} {Origin of biological
  homochirality by crystallization of an rna precursor on a magnetic
  surface},}\ }\href {\doibase 10.1126/sciadv.adg8274} {\bibfield  {journal}
  {\bibinfo  {journal} {Science Advances}\ }\textbf {\bibinfo {volume} {9}},\
  \bibinfo {pages} {eadg8274} (\bibinfo {year} {2023}{\natexlab{a}})},\ \Eprint
  {http://arxiv.org/abs/https://www.science.org/doi/pdf/10.1126/sciadv.adg8274}
  {https://www.science.org/doi/pdf/10.1126/sciadv.adg8274} \BibitemShut
  {NoStop}%
\bibitem [{\citenamefont {Ozturk}\ \emph
  {et~al.}(2023{\natexlab{b}})\citenamefont {Ozturk}, \citenamefont {Bhowmick},
  \citenamefont {Kapon}, \citenamefont {Sang}, \citenamefont {Kumar},
  \citenamefont {Paltiel}, \citenamefont {Naaman},\ and\ \citenamefont
  {Sasselov}}]{NatComms.14.6351}%
  \BibitemOpen
  \bibfield  {author} {\bibinfo {author} {\bibfnamefont {S.~F.}\ \bibnamefont
  {Ozturk}}, \bibinfo {author} {\bibfnamefont {D.~K.}\ \bibnamefont
  {Bhowmick}}, \bibinfo {author} {\bibfnamefont {Y.}~\bibnamefont {Kapon}},
  \bibinfo {author} {\bibfnamefont {Y.}~\bibnamefont {Sang}}, \bibinfo {author}
  {\bibfnamefont {A.}~\bibnamefont {Kumar}}, \bibinfo {author} {\bibfnamefont
  {Y.}~\bibnamefont {Paltiel}}, \bibinfo {author} {\bibfnamefont
  {R.}~\bibnamefont {Naaman}}, \ and\ \bibinfo {author} {\bibfnamefont {D.~D.}\
  \bibnamefont {Sasselov}},\ }\bibfield  {title} {\enquote {\bibinfo {title}
  {Chirality-induced avalanche magnetization of magnetite by an rna
  precursor},}\ }\href {\doibase 10.1038/s41467-023-42130-8} {\bibfield
  {journal} {\bibinfo  {journal} {Nature Communications}\ }\textbf {\bibinfo
  {volume} {14}},\ \bibinfo {pages} {6351} (\bibinfo {year}
  {2023}{\natexlab{b}})}\BibitemShut {NoStop}%
\bibitem [{\citenamefont {Kapon}\ \emph {et~al.}(2024)\citenamefont {Kapon},
  \citenamefont {Brann}, \citenamefont {Yochelis}, \citenamefont {Fransson},
  \citenamefont {Sasselov}, \citenamefont {Paltiel},\ and\ \citenamefont
  {Ozturk}}]{arXiv:2412.05720}%
  \BibitemOpen
  \bibfield  {author} {\bibinfo {author} {\bibfnamefont {Y.}~\bibnamefont
  {Kapon}}, \bibinfo {author} {\bibfnamefont {L.}~\bibnamefont {Brann}},
  \bibinfo {author} {\bibfnamefont {S.}~\bibnamefont {Yochelis}}, \bibinfo
  {author} {\bibfnamefont {J.}~\bibnamefont {Fransson}}, \bibinfo {author}
  {\bibfnamefont {D.~D.}\ \bibnamefont {Sasselov}}, \bibinfo {author}
  {\bibfnamefont {Y.}~\bibnamefont {Paltiel}}, \ and\ \bibinfo {author}
  {\bibfnamefont {S.~F.}\ \bibnamefont {Ozturk}},\ }\href
  {https://arxiv.org/abs/2412.05720} {\enquote {\bibinfo {title} {Non-classical
  temperature dependence of chirality-induced magnetization and its
  implications for rna's homochirality},}\ } (\bibinfo {year} {2024}),\ \Eprint
  {http://arxiv.org/abs/2412.05720} {arXiv:2412.05720 [physics.chem-ph]}
  \BibitemShut {NoStop}%
\bibitem [{\citenamefont {Ghosh}\ \emph {et~al.}(2019)\citenamefont {Ghosh},
  \citenamefont {Zhang}, \citenamefont {Tassinari}, \citenamefont {Mastai},
  \citenamefont {Lidor-Shalev}, \citenamefont {Naaman}, \citenamefont
  {M{\"o}llers}, \citenamefont {N{\"u}renberg}, \citenamefont {Zacharias},
  \citenamefont {Wei}, \citenamefont {Wierzbinski},\ and\ \citenamefont
  {Waldeck}}]{JPhysChemC.123.3024}%
  \BibitemOpen
  \bibfield  {author} {\bibinfo {author} {\bibfnamefont {K.~B.}\ \bibnamefont
  {Ghosh}}, \bibinfo {author} {\bibfnamefont {W.}~\bibnamefont {Zhang}},
  \bibinfo {author} {\bibfnamefont {F.}~\bibnamefont {Tassinari}}, \bibinfo
  {author} {\bibfnamefont {Y.}~\bibnamefont {Mastai}}, \bibinfo {author}
  {\bibfnamefont {O.}~\bibnamefont {Lidor-Shalev}}, \bibinfo {author}
  {\bibfnamefont {R.}~\bibnamefont {Naaman}}, \bibinfo {author} {\bibfnamefont
  {P.}~\bibnamefont {M{\"o}llers}}, \bibinfo {author} {\bibfnamefont
  {D.}~\bibnamefont {N{\"u}renberg}}, \bibinfo {author} {\bibfnamefont
  {H.}~\bibnamefont {Zacharias}}, \bibinfo {author} {\bibfnamefont
  {J.}~\bibnamefont {Wei}}, \bibinfo {author} {\bibfnamefont {E.}~\bibnamefont
  {Wierzbinski}}, \ and\ \bibinfo {author} {\bibfnamefont {D.~H.}\ \bibnamefont
  {Waldeck}},\ }\bibfield  {title} {\enquote {\bibinfo {title} {Controlling
  chemical selectivity in electrocatalysis with chiral cuo-coated
  electrodes},}\ }\href {\doibase 10.1021/acs.jpcc.8b12027} {\bibfield
  {journal} {\bibinfo  {journal} {The Journal of Physical Chemistry C}\
  }\textbf {\bibinfo {volume} {123}},\ \bibinfo {pages} {3024--3031} (\bibinfo
  {year} {2019})}\BibitemShut {NoStop}%
\bibitem [{\citenamefont {Ghosh}\ \emph
  {et~al.}(2020{\natexlab{a}})\citenamefont {Ghosh}, \citenamefont {Bloom},
  \citenamefont {Lu}, \citenamefont {Lamont},\ and\ \citenamefont
  {Waldeck}}]{JPhysChemC.124.22610}%
  \BibitemOpen
  \bibfield  {author} {\bibinfo {author} {\bibfnamefont {S.}~\bibnamefont
  {Ghosh}}, \bibinfo {author} {\bibfnamefont {B.~P.}\ \bibnamefont {Bloom}},
  \bibinfo {author} {\bibfnamefont {Y.}~\bibnamefont {Lu}}, \bibinfo {author}
  {\bibfnamefont {D.}~\bibnamefont {Lamont}}, \ and\ \bibinfo {author}
  {\bibfnamefont {D.~H.}\ \bibnamefont {Waldeck}},\ }\bibfield  {title}
  {\enquote {\bibinfo {title} {Increasing the efficiency of water splitting
  through spin polarization using cobalt oxide thin film catalysts},}\ }\href
  {\doibase 10.1021/acs.jpcc.0c07372} {\bibfield  {journal} {\bibinfo
  {journal} {The Journal of Physical Chemistry C}\ }\textbf {\bibinfo {volume}
  {124}},\ \bibinfo {pages} {22610--22618} (\bibinfo {year}
  {2020}{\natexlab{a}})}\BibitemShut {NoStop}%
\bibitem [{\citenamefont {Sang}\ \emph {et~al.}(2022)\citenamefont {Sang},
  \citenamefont {Tassinari}, \citenamefont {Santra}, \citenamefont {Zhang},
  \citenamefont {Fontanesi}, \citenamefont {Bloom}, \citenamefont {Waldeck},
  \citenamefont {Fransson},\ and\ \citenamefont {Naaman}}]{PNAS.30.2202650119}%
  \BibitemOpen
  \bibfield  {author} {\bibinfo {author} {\bibfnamefont {Y.}~\bibnamefont
  {Sang}}, \bibinfo {author} {\bibfnamefont {F.}~\bibnamefont {Tassinari}},
  \bibinfo {author} {\bibfnamefont {K.}~\bibnamefont {Santra}}, \bibinfo
  {author} {\bibfnamefont {W.}~\bibnamefont {Zhang}}, \bibinfo {author}
  {\bibfnamefont {C.}~\bibnamefont {Fontanesi}}, \bibinfo {author}
  {\bibfnamefont {B.~P.}\ \bibnamefont {Bloom}}, \bibinfo {author}
  {\bibfnamefont {D.~H.}\ \bibnamefont {Waldeck}}, \bibinfo {author}
  {\bibfnamefont {J.}~\bibnamefont {Fransson}}, \ and\ \bibinfo {author}
  {\bibfnamefont {R.}~\bibnamefont {Naaman}},\ }\bibfield  {title} {\enquote
  {\bibinfo {title} {Chirality enhances oxygen reduction},}\ }\href {\doibase
  10.1073/pnas.2202650119} {\bibfield  {journal} {\bibinfo  {journal}
  {Proceedings of the National Academy of Sciences}\ }\textbf {\bibinfo
  {volume} {119}},\ \bibinfo {pages} {e2202650119} (\bibinfo {year}
  {2022})}\BibitemShut {NoStop}%
\bibitem [{\citenamefont {Gupta}\ \emph
  {et~al.}(2023{\natexlab{a}})\citenamefont {Gupta}, \citenamefont {Sang},
  \citenamefont {Fontanesi}, \citenamefont {Turin},\ and\ \citenamefont
  {Naaman}}]{JPhysChemLett.14.1756}%
  \BibitemOpen
  \bibfield  {author} {\bibinfo {author} {\bibfnamefont {A.}~\bibnamefont
  {Gupta}}, \bibinfo {author} {\bibfnamefont {Y.}~\bibnamefont {Sang}},
  \bibinfo {author} {\bibfnamefont {C.}~\bibnamefont {Fontanesi}}, \bibinfo
  {author} {\bibfnamefont {L.}~\bibnamefont {Turin}}, \ and\ \bibinfo {author}
  {\bibfnamefont {R.}~\bibnamefont {Naaman}},\ }\bibfield  {title} {\enquote
  {\bibinfo {title} {Effect of anesthesia gases on the oxygen reduction
  reaction},}\ }\href {\doibase 10.1021/acs.jpclett.2c03753} {\bibfield
  {journal} {\bibinfo  {journal} {The Journal of Physical Chemistry Letters}\
  }\textbf {\bibinfo {volume} {14}},\ \bibinfo {pages} {1756--1761} (\bibinfo
  {year} {2023}{\natexlab{a}})}\BibitemShut {NoStop}%
\bibitem [{\citenamefont {Ben~Dor}\ \emph {et~al.}(2017)\citenamefont
  {Ben~Dor}, \citenamefont {Yochelis}, \citenamefont {Radko}, \citenamefont
  {Vankayala}, \citenamefont {Capua}, \citenamefont {Capua}, \citenamefont
  {Yang}, \citenamefont {Baczewski}, \citenamefont {Parkin}, \citenamefont
  {Naaman},\ and\ \citenamefont {Paltiel}}]{NatComms.8.14567}%
  \BibitemOpen
  \bibfield  {author} {\bibinfo {author} {\bibfnamefont {O.}~\bibnamefont
  {Ben~Dor}}, \bibinfo {author} {\bibfnamefont {S.}~\bibnamefont {Yochelis}},
  \bibinfo {author} {\bibfnamefont {A.}~\bibnamefont {Radko}}, \bibinfo
  {author} {\bibfnamefont {K.}~\bibnamefont {Vankayala}}, \bibinfo {author}
  {\bibfnamefont {E.}~\bibnamefont {Capua}}, \bibinfo {author} {\bibfnamefont
  {A.}~\bibnamefont {Capua}}, \bibinfo {author} {\bibfnamefont {S.-H.}\
  \bibnamefont {Yang}}, \bibinfo {author} {\bibfnamefont {L.~T.}\ \bibnamefont
  {Baczewski}}, \bibinfo {author} {\bibfnamefont {S.~S.~P.}\ \bibnamefont
  {Parkin}}, \bibinfo {author} {\bibfnamefont {R.}~\bibnamefont {Naaman}}, \
  and\ \bibinfo {author} {\bibfnamefont {Y.}~\bibnamefont {Paltiel}},\
  }\bibfield  {title} {\enquote {\bibinfo {title} {Magnetization switching in
  ferromagnets by adsorbed chiral molecules without current or external
  magnetic field},}\ }\href {\doibase 10.1038/ncomms14567} {\bibfield
  {journal} {\bibinfo  {journal} {Nature Communications}\ }\textbf {\bibinfo
  {volume} {8}},\ \bibinfo {pages} {14567} (\bibinfo {year}
  {2017})}\BibitemShut {NoStop}%
\bibitem [{\citenamefont {Alpern}\ \emph {et~al.}(2019)\citenamefont {Alpern},
  \citenamefont {Yavilberg}, \citenamefont {Dvir}, \citenamefont {Sukenik},
  \citenamefont {Klang}, \citenamefont {Yochelis}, \citenamefont {Cohen},
  \citenamefont {Grosfeld}, \citenamefont {Steinberg}, \citenamefont
  {Paltiel},\ and\ \citenamefont {Millo}}]{NanoLett.19.5167}%
  \BibitemOpen
  \bibfield  {author} {\bibinfo {author} {\bibfnamefont {H.}~\bibnamefont
  {Alpern}}, \bibinfo {author} {\bibfnamefont {K.}~\bibnamefont {Yavilberg}},
  \bibinfo {author} {\bibfnamefont {T.}~\bibnamefont {Dvir}}, \bibinfo {author}
  {\bibfnamefont {N.}~\bibnamefont {Sukenik}}, \bibinfo {author} {\bibfnamefont
  {M.}~\bibnamefont {Klang}}, \bibinfo {author} {\bibfnamefont
  {S.}~\bibnamefont {Yochelis}}, \bibinfo {author} {\bibfnamefont
  {H.}~\bibnamefont {Cohen}}, \bibinfo {author} {\bibfnamefont
  {E.}~\bibnamefont {Grosfeld}}, \bibinfo {author} {\bibfnamefont
  {H.}~\bibnamefont {Steinberg}}, \bibinfo {author} {\bibfnamefont
  {Y.}~\bibnamefont {Paltiel}}, \ and\ \bibinfo {author} {\bibfnamefont
  {O.}~\bibnamefont {Millo}},\ }\bibfield  {title} {\enquote {\bibinfo {title}
  {Magnetic-related states and order parameter induced in a conventional
  superconductor by nonmagnetic chiral molecules},}\ }\href {\doibase
  10.1021/acs.nanolett.9b01552} {\bibfield  {journal} {\bibinfo  {journal}
  {Nano Letters}\ }\textbf {\bibinfo {volume} {19}},\ \bibinfo {pages}
  {5167--5175} (\bibinfo {year} {2019})}\BibitemShut {NoStop}%
\bibitem [{\citenamefont {Fransson}\ and\ \citenamefont
  {Turin}(2024)}]{JPhysChemLett.15.6370}%
  \BibitemOpen
  \bibfield  {author} {\bibinfo {author} {\bibfnamefont {J.}~\bibnamefont
  {Fransson}}\ and\ \bibinfo {author} {\bibfnamefont {L.}~\bibnamefont
  {Turin}},\ }\bibfield  {title} {\enquote {\bibinfo {title} {Current induced
  spin-polarization in chiral molecules},}\ }\href {\doibase
  10.1021/acs.jpclett.4c01362} {\bibfield  {journal} {\bibinfo  {journal} {The
  Journal of Physical Chemistry Letters}\ }\textbf {\bibinfo {volume} {15}},\
  \bibinfo {pages} {6370--6374} (\bibinfo {year} {2024})}\BibitemShut {NoStop}%
\bibitem [{\citenamefont {Fransson}\ \emph {et~al.}(2025)\citenamefont
  {Fransson}, \citenamefont {Kapon}, \citenamefont {Brann}, \citenamefont
  {Yochelis}, \citenamefont {Sasselov}, \citenamefont {Paltiel},\ and\
  \citenamefont {Ozturk}}]{JPhysChemLett.16.2001}%
  \BibitemOpen
  \bibfield  {author} {\bibinfo {author} {\bibfnamefont {J.}~\bibnamefont
  {Fransson}}, \bibinfo {author} {\bibfnamefont {Y.}~\bibnamefont {Kapon}},
  \bibinfo {author} {\bibfnamefont {L.}~\bibnamefont {Brann}}, \bibinfo
  {author} {\bibfnamefont {S.}~\bibnamefont {Yochelis}}, \bibinfo {author}
  {\bibfnamefont {D.~D.}\ \bibnamefont {Sasselov}}, \bibinfo {author}
  {\bibfnamefont {Y.}~\bibnamefont {Paltiel}}, \ and\ \bibinfo {author}
  {\bibfnamefont {S.~F.}\ \bibnamefont {Ozturk}},\ }\bibfield  {title}
  {\enquote {\bibinfo {title} {Chiral phonons enhance ferromagnetism},}\ }\href
  {\doibase 10.1021/acs.jpclett.5c00304} {\bibfield  {journal} {\bibinfo
  {journal} {The Journal of Physical Chemistry Letters}\ }\textbf {\bibinfo
  {volume} {16}},\ \bibinfo {pages} {2001--2007} (\bibinfo {year}
  {2025})}\BibitemShut {NoStop}%
\bibitem [{\citenamefont {Ray}\ \emph {et~al.}(1999)\citenamefont {Ray},
  \citenamefont {Ananthavel}, \citenamefont {Waldeck},\ and\ \citenamefont
  {Naaman}}]{Science.283.814}%
  \BibitemOpen
  \bibfield  {author} {\bibinfo {author} {\bibfnamefont {K.}~\bibnamefont
  {Ray}}, \bibinfo {author} {\bibfnamefont {S.~P.}\ \bibnamefont {Ananthavel}},
  \bibinfo {author} {\bibfnamefont {D.~H.}\ \bibnamefont {Waldeck}}, \ and\
  \bibinfo {author} {\bibfnamefont {R.}~\bibnamefont {Naaman}},\ }\bibfield
  {title} {\enquote {\bibinfo {title} {Asymmetric scattering of polarized
  electrons by organized organic films of chiral molecules},}\ }\href {\doibase
  10.1126/science.283.5403.814} {\bibfield  {journal} {\bibinfo  {journal}
  {Science}\ }\textbf {\bibinfo {volume} {283}},\ \bibinfo {pages} {814--816}
  (\bibinfo {year} {1999})}\BibitemShut {NoStop}%
\bibitem [{\citenamefont {G{\"o}hler}\ \emph {et~al.}(2011)\citenamefont
  {G{\"o}hler}, \citenamefont {Hamelbeck}, \citenamefont {Markus},
  \citenamefont {Kettner}, \citenamefont {Hanne}, \citenamefont {Vager},
  \citenamefont {Naaman},\ and\ \citenamefont {Zacharias}}]{Science.331.894}%
  \BibitemOpen
  \bibfield  {author} {\bibinfo {author} {\bibfnamefont {B.}~\bibnamefont
  {G{\"o}hler}}, \bibinfo {author} {\bibfnamefont {V.}~\bibnamefont
  {Hamelbeck}}, \bibinfo {author} {\bibfnamefont {T.~Z.}\ \bibnamefont
  {Markus}}, \bibinfo {author} {\bibfnamefont {M.}~\bibnamefont {Kettner}},
  \bibinfo {author} {\bibfnamefont {G.~F.}\ \bibnamefont {Hanne}}, \bibinfo
  {author} {\bibfnamefont {Z.}~\bibnamefont {Vager}}, \bibinfo {author}
  {\bibfnamefont {R.}~\bibnamefont {Naaman}}, \ and\ \bibinfo {author}
  {\bibfnamefont {H.}~\bibnamefont {Zacharias}},\ }\bibfield  {title} {\enquote
  {\bibinfo {title} {Spin selectivity in electron transmission through
  self-assembled monolayers of double-stranded dna},}\ }\href {\doibase
  10.1126/science.1199339} {\bibfield  {journal} {\bibinfo  {journal}
  {Science}\ }\textbf {\bibinfo {volume} {331}},\ \bibinfo {pages} {894--897}
  (\bibinfo {year} {2011})}\BibitemShut {NoStop}%
\bibitem [{\citenamefont {Yeganeh}\ \emph {et~al.}(2009)\citenamefont
  {Yeganeh}, \citenamefont {Ratner}, \citenamefont {Medina},\ and\
  \citenamefont {Mujica}}]{JChemPhys.131.014707}%
  \BibitemOpen
  \bibfield  {author} {\bibinfo {author} {\bibfnamefont {S.}~\bibnamefont
  {Yeganeh}}, \bibinfo {author} {\bibfnamefont {M.~A.}\ \bibnamefont {Ratner}},
  \bibinfo {author} {\bibfnamefont {E.}~\bibnamefont {Medina}}, \ and\ \bibinfo
  {author} {\bibfnamefont {V.}~\bibnamefont {Mujica}},\ }\bibfield  {title}
  {\enquote {\bibinfo {title} {{Chiral electron transport: Scattering through
  helical potentials}},}\ }\href {\doibase 10.1063/1.3167404} {\bibfield
  {journal} {\bibinfo  {journal} {The Journal of Chemical Physics}\ }\textbf
  {\bibinfo {volume} {131}},\ \bibinfo {pages} {014707} (\bibinfo {year}
  {2009})},\ \Eprint
  {http://arxiv.org/abs/https://pubs.aip.org/aip/jcp/article-pdf/doi/10.1063/1.3167404/16105082/014707\_1\_online.pdf}
  {https://pubs.aip.org/aip/jcp/article-pdf/doi/10.1063/1.3167404/16105082/014707\_1\_online.pdf}
  \BibitemShut {NoStop}%
\bibitem [{\citenamefont {Medina}\ \emph {et~al.}(2012)\citenamefont {Medina},
  \citenamefont {López}, \citenamefont {Ratner},\ and\ \citenamefont
  {Mujica}}]{EPL.99.17006}%
  \BibitemOpen
  \bibfield  {author} {\bibinfo {author} {\bibfnamefont {E.}~\bibnamefont
  {Medina}}, \bibinfo {author} {\bibfnamefont {F.}~\bibnamefont {López}},
  \bibinfo {author} {\bibfnamefont {M.~A.}\ \bibnamefont {Ratner}}, \ and\
  \bibinfo {author} {\bibfnamefont {V.}~\bibnamefont {Mujica}},\ }\bibfield
  {title} {\enquote {\bibinfo {title} {Chiral molecular films as electron
  polarizers and polarization modulators},}\ }\href {\doibase
  10.1209/0295-5075/99/17006} {\bibfield  {journal} {\bibinfo  {journal}
  {Europhysics Letters}\ }\textbf {\bibinfo {volume} {99}},\ \bibinfo {pages}
  {17006} (\bibinfo {year} {2012})}\BibitemShut {NoStop}%
\bibitem [{\citenamefont {Varela}\ \emph {et~al.}(2013)\citenamefont {Varela},
  \citenamefont {Medina}, \citenamefont {López},\ and\ \citenamefont
  {Mujica}}]{JPCM.26.015008}%
  \BibitemOpen
  \bibfield  {author} {\bibinfo {author} {\bibfnamefont {S.}~\bibnamefont
  {Varela}}, \bibinfo {author} {\bibfnamefont {E.}~\bibnamefont {Medina}},
  \bibinfo {author} {\bibfnamefont {F.}~\bibnamefont {López}}, \ and\ \bibinfo
  {author} {\bibfnamefont {V.}~\bibnamefont {Mujica}},\ }\bibfield  {title}
  {\enquote {\bibinfo {title} {Inelastic electron scattering from a helical
  potential: transverse polarization and the structure factor in the single
  scattering approximation},}\ }\href {\doibase 10.1088/0953-8984/26/1/015008}
  {\bibfield  {journal} {\bibinfo  {journal} {Journal of Physics: Condensed
  Matter}\ }\textbf {\bibinfo {volume} {26}},\ \bibinfo {pages} {015008}
  (\bibinfo {year} {2013})}\BibitemShut {NoStop}%
\bibitem [{\citenamefont {Eremko}\ and\ \citenamefont
  {Loktev}(2013)}]{PhysRevB.88.165409}%
  \BibitemOpen
  \bibfield  {author} {\bibinfo {author} {\bibfnamefont {A.~A.}\ \bibnamefont
  {Eremko}}\ and\ \bibinfo {author} {\bibfnamefont {V.~M.}\ \bibnamefont
  {Loktev}},\ }\bibfield  {title} {\enquote {\bibinfo {title} {Spin sensitive
  electron transmission through helical potentials},}\ }\href {\doibase
  10.1103/PhysRevB.88.165409} {\bibfield  {journal} {\bibinfo  {journal} {Phys.
  Rev. B}\ }\textbf {\bibinfo {volume} {88}},\ \bibinfo {pages} {165409}
  (\bibinfo {year} {2013})}\BibitemShut {NoStop}%
\bibitem [{\citenamefont {Medina}\ \emph {et~al.}(2015)\citenamefont {Medina},
  \citenamefont {Gonz{\'a}lez-Arraga}, \citenamefont {Finkelstein-Shapiro},
  \citenamefont {Berche},\ and\ \citenamefont {Mujica}}]{JChemPhys.142.194308}%
  \BibitemOpen
  \bibfield  {author} {\bibinfo {author} {\bibfnamefont {E.}~\bibnamefont
  {Medina}}, \bibinfo {author} {\bibfnamefont {L.~A.}\ \bibnamefont
  {Gonz{\'a}lez-Arraga}}, \bibinfo {author} {\bibfnamefont {D.}~\bibnamefont
  {Finkelstein-Shapiro}}, \bibinfo {author} {\bibfnamefont {B.}~\bibnamefont
  {Berche}}, \ and\ \bibinfo {author} {\bibfnamefont {V.}~\bibnamefont
  {Mujica}},\ }\bibfield  {title} {\enquote {\bibinfo {title} {{Continuum model
  for chiral induced spin selectivity in helical molecules}},}\ }\href
  {\doibase 10.1063/1.4921310} {\bibfield  {journal} {\bibinfo  {journal} {The
  Journal of Chemical Physics}\ }\textbf {\bibinfo {volume} {142}},\ \bibinfo
  {pages} {194308} (\bibinfo {year} {2015})},\ \Eprint
  {http://arxiv.org/abs/https://pubs.aip.org/aip/jcp/article-pdf/doi/10.1063/1.4921310/13233894/194308\_1\_online.pdf}
  {https://pubs.aip.org/aip/jcp/article-pdf/doi/10.1063/1.4921310/13233894/194308\_1\_online.pdf}
  \BibitemShut {NoStop}%
\bibitem [{\citenamefont {D\'{\i}az}\ \emph {et~al.}(2018)\citenamefont
  {D\'{\i}az}, \citenamefont {Contreras}, \citenamefont {Hern\'andez},\ and\
  \citenamefont {Dom\'{\i}nguez-Adame}}]{PhysRevE.98.052221}%
  \BibitemOpen
  \bibfield  {author} {\bibinfo {author} {\bibfnamefont {E.}~\bibnamefont
  {D\'{\i}az}}, \bibinfo {author} {\bibfnamefont {A.}~\bibnamefont
  {Contreras}}, \bibinfo {author} {\bibfnamefont {J.}~\bibnamefont
  {Hern\'andez}}, \ and\ \bibinfo {author} {\bibfnamefont {F.}~\bibnamefont
  {Dom\'{\i}nguez-Adame}},\ }\bibfield  {title} {\enquote {\bibinfo {title}
  {Effective nonlinear model for electron transport in deformable helical
  molecules},}\ }\href {\doibase 10.1103/PhysRevE.98.052221} {\bibfield
  {journal} {\bibinfo  {journal} {Phys. Rev. E}\ }\textbf {\bibinfo {volume}
  {98}},\ \bibinfo {pages} {052221} (\bibinfo {year} {2018})}\BibitemShut
  {NoStop}%
\bibitem [{\citenamefont {Yang}, \citenamefont {van~der Wal},\ and\
  \citenamefont {van Wees}(2019)}]{PhysRevB.99.024418}%
  \BibitemOpen
  \bibfield  {author} {\bibinfo {author} {\bibfnamefont {X.}~\bibnamefont
  {Yang}}, \bibinfo {author} {\bibfnamefont {C.~H.}\ \bibnamefont {van~der
  Wal}}, \ and\ \bibinfo {author} {\bibfnamefont {B.~J.}\ \bibnamefont {van
  Wees}},\ }\bibfield  {title} {\enquote {\bibinfo {title} {Spin-dependent
  electron transmission model for chiral molecules in mesoscopic devices},}\
  }\href {\doibase 10.1103/PhysRevB.99.024418} {\bibfield  {journal} {\bibinfo
  {journal} {Phys. Rev. B}\ }\textbf {\bibinfo {volume} {99}},\ \bibinfo
  {pages} {024418} (\bibinfo {year} {2019})}\BibitemShut {NoStop}%
\bibitem [{\citenamefont {Díaz}\ \emph
  {et~al.}(2018{\natexlab{a}})\citenamefont {Díaz}, \citenamefont {Albares},
  \citenamefont {Estévez}, \citenamefont {Cerveró}, \citenamefont {Gaul},
  \citenamefont {Diez},\ and\ \citenamefont
  {Domínguez-Adame}}]{NJP.20.043055}%
  \BibitemOpen
  \bibfield  {author} {\bibinfo {author} {\bibfnamefont {E.}~\bibnamefont
  {Díaz}}, \bibinfo {author} {\bibfnamefont {P.}~\bibnamefont {Albares}},
  \bibinfo {author} {\bibfnamefont {P.~G.}\ \bibnamefont {Estévez}}, \bibinfo
  {author} {\bibfnamefont {J.~M.}\ \bibnamefont {Cerveró}}, \bibinfo {author}
  {\bibfnamefont {C.}~\bibnamefont {Gaul}}, \bibinfo {author} {\bibfnamefont
  {E.}~\bibnamefont {Diez}}, \ and\ \bibinfo {author} {\bibfnamefont
  {F.}~\bibnamefont {Domínguez-Adame}},\ }\bibfield  {title} {\enquote
  {\bibinfo {title} {Spin dynamics in helical molecules with nonlinear
  interactions},}\ }\href {\doibase 10.1088/1367-2630/aabb91} {\bibfield
  {journal} {\bibinfo  {journal} {New Journal of Physics}\ }\textbf {\bibinfo
  {volume} {20}},\ \bibinfo {pages} {043055} (\bibinfo {year}
  {2018}{\natexlab{a}})}\BibitemShut {NoStop}%
\bibitem [{\citenamefont {Michaeli}\ and\ \citenamefont
  {Naaman}(2019)}]{JPhysChemC.123.17043}%
  \BibitemOpen
  \bibfield  {author} {\bibinfo {author} {\bibfnamefont {K.}~\bibnamefont
  {Michaeli}}\ and\ \bibinfo {author} {\bibfnamefont {R.}~\bibnamefont
  {Naaman}},\ }\bibfield  {title} {\enquote {\bibinfo {title} {Origin of
  spin-dependent tunneling through chiral molecules},}\ }\href {\doibase
  10.1021/acs.jpcc.9b05020} {\bibfield  {journal} {\bibinfo  {journal} {The
  Journal of Physical Chemistry C}\ }\textbf {\bibinfo {volume} {123}},\
  \bibinfo {pages} {17043--17048} (\bibinfo {year} {2019})},\ \Eprint
  {http://arxiv.org/abs/https://doi.org/10.1021/acs.jpcc.9b05020}
  {https://doi.org/10.1021/acs.jpcc.9b05020} \BibitemShut {NoStop}%
\bibitem [{\citenamefont {Gutierrez}\ \emph {et~al.}(2012)\citenamefont
  {Gutierrez}, \citenamefont {D\'{\i}az}, \citenamefont {Naaman},\ and\
  \citenamefont {Cuniberti}}]{PhysRevB.85.081404}%
  \BibitemOpen
  \bibfield  {author} {\bibinfo {author} {\bibfnamefont {R.}~\bibnamefont
  {Gutierrez}}, \bibinfo {author} {\bibfnamefont {E.}~\bibnamefont
  {D\'{\i}az}}, \bibinfo {author} {\bibfnamefont {R.}~\bibnamefont {Naaman}}, \
  and\ \bibinfo {author} {\bibfnamefont {G.}~\bibnamefont {Cuniberti}},\
  }\bibfield  {title} {\enquote {\bibinfo {title} {Spin-selective transport
  through helical molecular systems},}\ }\href {\doibase
  10.1103/PhysRevB.85.081404} {\bibfield  {journal} {\bibinfo  {journal} {Phys.
  Rev. B}\ }\textbf {\bibinfo {volume} {85}},\ \bibinfo {pages} {081404}
  (\bibinfo {year} {2012})}\BibitemShut {NoStop}%
\bibitem [{\citenamefont {Guo}\ and\ \citenamefont
  {Sun}(2012)}]{PhysRevLett.108.218102}%
  \BibitemOpen
  \bibfield  {author} {\bibinfo {author} {\bibfnamefont {A.-M.}\ \bibnamefont
  {Guo}}\ and\ \bibinfo {author} {\bibfnamefont {Q.-f.}\ \bibnamefont {Sun}},\
  }\bibfield  {title} {\enquote {\bibinfo {title} {Spin-selective transport of
  electrons in dna double helix},}\ }\href {\doibase
  10.1103/PhysRevLett.108.218102} {\bibfield  {journal} {\bibinfo  {journal}
  {Phys. Rev. Lett.}\ }\textbf {\bibinfo {volume} {108}},\ \bibinfo {pages}
  {218102} (\bibinfo {year} {2012})}\BibitemShut {NoStop}%
\bibitem [{\citenamefont {Guo}\ and\ \citenamefont
  {Sun}(2014)}]{PNAS.111.11658}%
  \BibitemOpen
  \bibfield  {author} {\bibinfo {author} {\bibfnamefont {A.-M.}\ \bibnamefont
  {Guo}}\ and\ \bibinfo {author} {\bibfnamefont {Q.-F.}\ \bibnamefont {Sun}},\
  }\bibfield  {title} {\enquote {\bibinfo {title} {Spin-dependent electron
  transport in protein-like single-helical molecules},}\ }\href {\doibase
  10.1073/pnas.1407716111} {\bibfield  {journal} {\bibinfo  {journal}
  {Proceedings of the National Academy of Sciences}\ }\textbf {\bibinfo
  {volume} {111}},\ \bibinfo {pages} {11658--11662} (\bibinfo {year} {2014})},\
  \Eprint
  {http://arxiv.org/abs/https://www.pnas.org/doi/pdf/10.1073/pnas.1407716111}
  {https://www.pnas.org/doi/pdf/10.1073/pnas.1407716111} \BibitemShut {NoStop}%
\bibitem [{\citenamefont {Rai}\ and\ \citenamefont
  {Galperin}(2013)}]{JPhysChemC.117.13730}%
  \BibitemOpen
  \bibfield  {author} {\bibinfo {author} {\bibfnamefont {D.}~\bibnamefont
  {Rai}}\ and\ \bibinfo {author} {\bibfnamefont {M.}~\bibnamefont {Galperin}},\
  }\bibfield  {title} {\enquote {\bibinfo {title} {Electrically driven spin
  currents in dna},}\ }\href {\doibase 10.1021/jp404066y} {\bibfield  {journal}
  {\bibinfo  {journal} {The Journal of Physical Chemistry C}\ }\textbf
  {\bibinfo {volume} {117}},\ \bibinfo {pages} {13730--13737} (\bibinfo {year}
  {2013})}\BibitemShut {NoStop}%
\bibitem [{\citenamefont {Matityahu}\ \emph {et~al.}(2016)\citenamefont
  {Matityahu}, \citenamefont {Utsumi}, \citenamefont {Aharony}, \citenamefont
  {Entin-Wohlman},\ and\ \citenamefont {Balseiro}}]{PhysRevB.93.075407}%
  \BibitemOpen
  \bibfield  {author} {\bibinfo {author} {\bibfnamefont {S.}~\bibnamefont
  {Matityahu}}, \bibinfo {author} {\bibfnamefont {Y.}~\bibnamefont {Utsumi}},
  \bibinfo {author} {\bibfnamefont {A.}~\bibnamefont {Aharony}}, \bibinfo
  {author} {\bibfnamefont {O.}~\bibnamefont {Entin-Wohlman}}, \ and\ \bibinfo
  {author} {\bibfnamefont {C.~A.}\ \bibnamefont {Balseiro}},\ }\bibfield
  {title} {\enquote {\bibinfo {title} {Spin-dependent transport through a
  chiral molecule in the presence of spin-orbit interaction and nonunitary
  effects},}\ }\href {\doibase 10.1103/PhysRevB.93.075407} {\bibfield
  {journal} {\bibinfo  {journal} {Phys. Rev. B}\ }\textbf {\bibinfo {volume}
  {93}},\ \bibinfo {pages} {075407} (\bibinfo {year} {2016})}\BibitemShut
  {NoStop}%
\bibitem [{\citenamefont {Varela}, \citenamefont {Mujica},\ and\ \citenamefont
  {Medina}(2016)}]{PhysRevB.93.155436}%
  \BibitemOpen
  \bibfield  {author} {\bibinfo {author} {\bibfnamefont {S.}~\bibnamefont
  {Varela}}, \bibinfo {author} {\bibfnamefont {V.}~\bibnamefont {Mujica}}, \
  and\ \bibinfo {author} {\bibfnamefont {E.}~\bibnamefont {Medina}},\
  }\bibfield  {title} {\enquote {\bibinfo {title} {Effective spin-orbit
  couplings in an analytical tight-binding model of dna: Spin filtering and
  chiral spin transport},}\ }\href {\doibase 10.1103/PhysRevB.93.155436}
  {\bibfield  {journal} {\bibinfo  {journal} {Phys. Rev. B}\ }\textbf {\bibinfo
  {volume} {93}},\ \bibinfo {pages} {155436} (\bibinfo {year}
  {2016})}\BibitemShut {NoStop}%
\bibitem [{\citenamefont {Behnia}, \citenamefont {Fathizadeh},\ and\
  \citenamefont {Akhshani}(2016)}]{ChemPhys.477.61}%
  \BibitemOpen
  \bibfield  {author} {\bibinfo {author} {\bibfnamefont {S.}~\bibnamefont
  {Behnia}}, \bibinfo {author} {\bibfnamefont {S.}~\bibnamefont {Fathizadeh}},
  \ and\ \bibinfo {author} {\bibfnamefont {A.}~\bibnamefont {Akhshani}},\
  }\bibfield  {title} {\enquote {\bibinfo {title} {Modeling spin selectivity in
  charge transfer across the dna/gold interface},}\ }\href {\doibase
  https://doi.org/10.1016/j.chemphys.2016.08.016} {\bibfield  {journal}
  {\bibinfo  {journal} {Chemical Physics}\ }\textbf {\bibinfo {volume} {477}},\
  \bibinfo {pages} {61--73} (\bibinfo {year} {2016})}\BibitemShut {NoStop}%
\bibitem [{\citenamefont {Maslyuk}\ \emph {et~al.}(2018)\citenamefont
  {Maslyuk}, \citenamefont {Gutierrez}, \citenamefont {Dianat}, \citenamefont
  {Mujica},\ and\ \citenamefont {Cuniberti}}]{JPhysChemLett.9.5453}%
  \BibitemOpen
  \bibfield  {author} {\bibinfo {author} {\bibfnamefont {V.~V.}\ \bibnamefont
  {Maslyuk}}, \bibinfo {author} {\bibfnamefont {R.}~\bibnamefont {Gutierrez}},
  \bibinfo {author} {\bibfnamefont {A.}~\bibnamefont {Dianat}}, \bibinfo
  {author} {\bibfnamefont {V.}~\bibnamefont {Mujica}}, \ and\ \bibinfo {author}
  {\bibfnamefont {G.}~\bibnamefont {Cuniberti}},\ }\bibfield  {title} {\enquote
  {\bibinfo {title} {Enhanced magnetoresistance in chiral molecular
  junctions},}\ }\href {\doibase 10.1021/acs.jpclett.8b02360} {\bibfield
  {journal} {\bibinfo  {journal} {The Journal of Physical Chemistry Letters}\
  }\textbf {\bibinfo {volume} {9}},\ \bibinfo {pages} {5453--5459} (\bibinfo
  {year} {2018})},\ \bibinfo {note} {pMID: 30188726},\ \Eprint
  {http://arxiv.org/abs/https://doi.org/10.1021/acs.jpclett.8b02360}
  {https://doi.org/10.1021/acs.jpclett.8b02360} \BibitemShut {NoStop}%
\bibitem [{\citenamefont {Díaz}\ \emph
  {et~al.}(2018{\natexlab{b}})\citenamefont {Díaz}, \citenamefont
  {Domínguez-Adame}, \citenamefont {Gutierrez}, \citenamefont {Cuniberti},\
  and\ \citenamefont {Mujica}}]{JPhysChemLett.9.5753}%
  \BibitemOpen
  \bibfield  {author} {\bibinfo {author} {\bibfnamefont {E.}~\bibnamefont
  {Díaz}}, \bibinfo {author} {\bibfnamefont {F.}~\bibnamefont
  {Domínguez-Adame}}, \bibinfo {author} {\bibfnamefont {R.}~\bibnamefont
  {Gutierrez}}, \bibinfo {author} {\bibfnamefont {G.}~\bibnamefont
  {Cuniberti}}, \ and\ \bibinfo {author} {\bibfnamefont {V.}~\bibnamefont
  {Mujica}},\ }\bibfield  {title} {\enquote {\bibinfo {title} {Thermal
  decoherence and disorder effects on chiral-induced spin selectivity},}\
  }\href {\doibase 10.1021/acs.jpclett.8b02196} {\bibfield  {journal} {\bibinfo
   {journal} {The Journal of Physical Chemistry Letters}\ }\textbf {\bibinfo
  {volume} {9}},\ \bibinfo {pages} {5753--5758} (\bibinfo {year}
  {2018}{\natexlab{b}})},\ \bibinfo {note} {pMID: 30212207},\ \Eprint
  {http://arxiv.org/abs/https://doi.org/10.1021/acs.jpclett.8b02196}
  {https://doi.org/10.1021/acs.jpclett.8b02196} \BibitemShut {NoStop}%
\bibitem [{\citenamefont {Zöllner}\ \emph {et~al.}(2020)\citenamefont
  {Zöllner}, \citenamefont {Varela}, \citenamefont {Medina}, \citenamefont
  {Mujica},\ and\ \citenamefont {Herrmann}}]{JChemTheoryComput.16.2914}%
  \BibitemOpen
  \bibfield  {author} {\bibinfo {author} {\bibfnamefont {M.~S.}\ \bibnamefont
  {Zöllner}}, \bibinfo {author} {\bibfnamefont {S.}~\bibnamefont {Varela}},
  \bibinfo {author} {\bibfnamefont {E.}~\bibnamefont {Medina}}, \bibinfo
  {author} {\bibfnamefont {V.}~\bibnamefont {Mujica}}, \ and\ \bibinfo {author}
  {\bibfnamefont {C.}~\bibnamefont {Herrmann}},\ }\bibfield  {title} {\enquote
  {\bibinfo {title} {Insight into the origin of chiral-induced spin selectivity
  from a symmetry analysis of electronic transmission},}\ }\href {\doibase
  10.1021/acs.jctc.9b01078} {\bibfield  {journal} {\bibinfo  {journal} {Journal
  of Chemical Theory and Computation}\ }\textbf {\bibinfo {volume} {16}},\
  \bibinfo {pages} {2914--2929} (\bibinfo {year} {2020})},\ \bibinfo {note}
  {pMID: 32271568},\ \Eprint
  {http://arxiv.org/abs/https://doi.org/10.1021/acs.jctc.9b01078}
  {https://doi.org/10.1021/acs.jctc.9b01078} \BibitemShut {NoStop}%
\bibitem [{\citenamefont {Ghazaryan}, \citenamefont {Lemeshko},\ and\
  \citenamefont {Volosniev}(2020)}]{CommunPhys.3.178}%
  \BibitemOpen
  \bibfield  {author} {\bibinfo {author} {\bibfnamefont {A.}~\bibnamefont
  {Ghazaryan}}, \bibinfo {author} {\bibfnamefont {M.}~\bibnamefont {Lemeshko}},
  \ and\ \bibinfo {author} {\bibfnamefont {A.~G.}\ \bibnamefont {Volosniev}},\
  }\bibfield  {title} {\enquote {\bibinfo {title} {Filtering spins by
  scattering from a lattice of point magnets},}\ }\href {\doibase
  10.1038/s42005-020-00445-8} {\bibfield  {journal} {\bibinfo  {journal}
  {Communications Physics}\ }\textbf {\bibinfo {volume} {3}},\ \bibinfo {pages}
  {178} (\bibinfo {year} {2020})}\BibitemShut {NoStop}%
\bibitem [{\citenamefont {Shitade}\ and\ \citenamefont
  {Minamitani}(2020)}]{NewJPhys.22.113023}%
  \BibitemOpen
  \bibfield  {author} {\bibinfo {author} {\bibfnamefont {A.}~\bibnamefont
  {Shitade}}\ and\ \bibinfo {author} {\bibfnamefont {E.}~\bibnamefont
  {Minamitani}},\ }\bibfield  {title} {\enquote {\bibinfo {title} {Geometric
  spin–orbit coupling and chirality-induced spin selectivity},}\ }\href
  {\doibase 10.1088/1367-2630/abc920} {\bibfield  {journal} {\bibinfo
  {journal} {New Journal of Physics}\ }\textbf {\bibinfo {volume} {22}},\
  \bibinfo {pages} {113023} (\bibinfo {year} {2020})}\BibitemShut {NoStop}%
\bibitem [{\citenamefont {Du}, \citenamefont {Fu},\ and\ \citenamefont
  {Wu}(2020)}]{PhysRevB.102.035431}%
  \BibitemOpen
  \bibfield  {author} {\bibinfo {author} {\bibfnamefont {G.-F.}\ \bibnamefont
  {Du}}, \bibinfo {author} {\bibfnamefont {H.-H.}\ \bibnamefont {Fu}}, \ and\
  \bibinfo {author} {\bibfnamefont {R.}~\bibnamefont {Wu}},\ }\bibfield
  {title} {\enquote {\bibinfo {title} {Vibration-enhanced spin-selective
  transport of electrons in the dna double helix},}\ }\href {\doibase
  10.1103/PhysRevB.102.035431} {\bibfield  {journal} {\bibinfo  {journal}
  {Phys. Rev. B}\ }\textbf {\bibinfo {volume} {102}},\ \bibinfo {pages}
  {035431} (\bibinfo {year} {2020})}\BibitemShut {NoStop}%
\bibitem [{\citenamefont {Fay}\ and\ \citenamefont
  {Limmer}(2021)}]{NanoLett.21.6696}%
  \BibitemOpen
  \bibfield  {author} {\bibinfo {author} {\bibfnamefont {T.~P.}\ \bibnamefont
  {Fay}}\ and\ \bibinfo {author} {\bibfnamefont {D.~T.}\ \bibnamefont
  {Limmer}},\ }\bibfield  {title} {\enquote {\bibinfo {title} {Origin of
  chirality induced spin selectivity in photoinduced electron transfer},}\
  }\href {\doibase 10.1021/acs.nanolett.1c02370} {\bibfield  {journal}
  {\bibinfo  {journal} {Nano Letters}\ }\textbf {\bibinfo {volume} {21}},\
  \bibinfo {pages} {6696--6702} (\bibinfo {year} {2021})},\ \bibinfo {note}
  {pMID: 34291928},\ \Eprint
  {http://arxiv.org/abs/https://doi.org/10.1021/acs.nanolett.1c02370}
  {https://doi.org/10.1021/acs.nanolett.1c02370} \BibitemShut {NoStop}%
\bibitem [{\citenamefont {Wang}\ \emph {et~al.}(2021)\citenamefont {Wang},
  \citenamefont {Guo}, \citenamefont {Sun},\ and\ \citenamefont
  {Yan}}]{JPhysChemLett.12.10262}%
  \BibitemOpen
  \bibfield  {author} {\bibinfo {author} {\bibfnamefont {C.}~\bibnamefont
  {Wang}}, \bibinfo {author} {\bibfnamefont {A.-M.}\ \bibnamefont {Guo}},
  \bibinfo {author} {\bibfnamefont {Q.-F.}\ \bibnamefont {Sun}}, \ and\
  \bibinfo {author} {\bibfnamefont {Y.}~\bibnamefont {Yan}},\ }\bibfield
  {title} {\enquote {\bibinfo {title} {Efficient spin-dependent charge
  transmission and improved enantioselective discrimination capability in
  self-assembled chiral coordinated monolayers},}\ }\href {\doibase
  10.1021/acs.jpclett.1c03106} {\bibfield  {journal} {\bibinfo  {journal} {The
  Journal of Physical Chemistry Letters}\ }\textbf {\bibinfo {volume} {12}},\
  \bibinfo {pages} {10262--10269} (\bibinfo {year} {2021})},\ \bibinfo {note}
  {pMID: 34652163},\ \Eprint
  {http://arxiv.org/abs/https://doi.org/10.1021/acs.jpclett.1c03106}
  {https://doi.org/10.1021/acs.jpclett.1c03106} \BibitemShut {NoStop}%
\bibitem [{\citenamefont {Wang}, \citenamefont {Mujica},\ and\ \citenamefont
  {Lai}(2021)}]{NanoLett.21.10423}%
  \BibitemOpen
  \bibfield  {author} {\bibinfo {author} {\bibfnamefont {C.-Z.}\ \bibnamefont
  {Wang}}, \bibinfo {author} {\bibfnamefont {V.}~\bibnamefont {Mujica}}, \ and\
  \bibinfo {author} {\bibfnamefont {Y.-C.}\ \bibnamefont {Lai}},\ }\bibfield
  {title} {\enquote {\bibinfo {title} {Spin fano resonances in chiral
  molecules: An alternative mechanism for the ciss effect and experimental
  implications},}\ }\href {\doibase 10.1021/acs.nanolett.1c03770} {\bibfield
  {journal} {\bibinfo  {journal} {Nano Letters}\ }\textbf {\bibinfo {volume}
  {21}},\ \bibinfo {pages} {10423--10430} (\bibinfo {year} {2021})},\ \bibinfo
  {note} {pMID: 34846905},\ \Eprint
  {http://arxiv.org/abs/https://doi.org/10.1021/acs.nanolett.1c03770}
  {https://doi.org/10.1021/acs.nanolett.1c03770} \BibitemShut {NoStop}%
\bibitem [{\citenamefont {Li}, \citenamefont {Nan},\ and\ \citenamefont
  {Pan}(2020)}]{PhysRevLett.125.263002}%
  \BibitemOpen
  \bibfield  {author} {\bibinfo {author} {\bibfnamefont {X.}~\bibnamefont
  {Li}}, \bibinfo {author} {\bibfnamefont {J.}~\bibnamefont {Nan}}, \ and\
  \bibinfo {author} {\bibfnamefont {X.}~\bibnamefont {Pan}},\ }\bibfield
  {title} {\enquote {\bibinfo {title} {Chiral induced spin selectivity as a
  spontaneous intertwined order},}\ }\href {\doibase
  10.1103/PhysRevLett.125.263002} {\bibfield  {journal} {\bibinfo  {journal}
  {Phys. Rev. Lett.}\ }\textbf {\bibinfo {volume} {125}},\ \bibinfo {pages}
  {263002} (\bibinfo {year} {2020})}\BibitemShut {NoStop}%
\bibitem [{\citenamefont {Chiesa}\ \emph {et~al.}(2024)\citenamefont {Chiesa},
  \citenamefont {Garlatti}, \citenamefont {Mezzadri}, \citenamefont {Celada},
  \citenamefont {Sessoli}, \citenamefont {Wasielewski}, \citenamefont {Bittl},
  \citenamefont {Santini},\ and\ \citenamefont {Carretta}}]{NanoLett.24.12133}%
  \BibitemOpen
  \bibfield  {author} {\bibinfo {author} {\bibfnamefont {A.}~\bibnamefont
  {Chiesa}}, \bibinfo {author} {\bibfnamefont {E.}~\bibnamefont {Garlatti}},
  \bibinfo {author} {\bibfnamefont {M.}~\bibnamefont {Mezzadri}}, \bibinfo
  {author} {\bibfnamefont {L.}~\bibnamefont {Celada}}, \bibinfo {author}
  {\bibfnamefont {R.}~\bibnamefont {Sessoli}}, \bibinfo {author} {\bibfnamefont
  {M.~R.}\ \bibnamefont {Wasielewski}}, \bibinfo {author} {\bibfnamefont
  {R.}~\bibnamefont {Bittl}}, \bibinfo {author} {\bibfnamefont
  {P.}~\bibnamefont {Santini}}, \ and\ \bibinfo {author} {\bibfnamefont
  {S.}~\bibnamefont {Carretta}},\ }\bibfield  {title} {\enquote {\bibinfo
  {title} {Many-body models for chirality-induced spin selectivity in electron
  transfer},}\ }\href {\doibase 10.1021/acs.nanolett.4c02912} {\bibfield
  {journal} {\bibinfo  {journal} {Nano Letters}\ }\textbf {\bibinfo {volume}
  {24}},\ \bibinfo {pages} {12133--12139} (\bibinfo {year} {2024})}\BibitemShut
  {NoStop}%
\bibitem [{\citenamefont {Bardarson}(2008)}]{JPAMT.41.405203}%
  \BibitemOpen
  \bibfield  {author} {\bibinfo {author} {\bibfnamefont {J.~H.}\ \bibnamefont
  {Bardarson}},\ }\bibfield  {title} {\enquote {\bibinfo {title} {A proof of
  the kramers degeneracy of transmission eigenvalues from antisymmetry of the
  scattering matrix},}\ }\href {\doibase 10.1088/1751-8113/41/40/405203}
  {\bibfield  {journal} {\bibinfo  {journal} {Journal of Physics A:
  Mathematical and Theoretical}\ }\textbf {\bibinfo {volume} {41}},\ \bibinfo
  {pages} {405203} (\bibinfo {year} {2008})}\BibitemShut {NoStop}%
\bibitem [{\citenamefont {Fradkin}(2013)}]{Fradkin2013}%
  \BibitemOpen
  \bibfield  {author} {\bibinfo {author} {\bibfnamefont {E.}~\bibnamefont
  {Fradkin}},\ }\href {\doibase DOI: 10.1017/CBO9781139015509} {\emph {\bibinfo
  {title} {Field Theories of Condensed Matter Physics}}},\ \bibinfo {edition}
  {2nd}\ ed.\ (\bibinfo  {publisher} {Cambridge University Press},\ \bibinfo
  {address} {Cambridge},\ \bibinfo {year} {2013})\BibitemShut {NoStop}%
\bibitem [{\citenamefont {Ghosh}\ \emph
  {et~al.}(2020{\natexlab{b}})\citenamefont {Ghosh}, \citenamefont {Mishra},
  \citenamefont {Avigad}, \citenamefont {Bloom}, \citenamefont {Baczewski},
  \citenamefont {Yochelis}, \citenamefont {Paltiel}, \citenamefont {Naaman},\
  and\ \citenamefont {Waldeck}}]{JPhysChemLett.11.1550}%
  \BibitemOpen
  \bibfield  {author} {\bibinfo {author} {\bibfnamefont {S.}~\bibnamefont
  {Ghosh}}, \bibinfo {author} {\bibfnamefont {S.}~\bibnamefont {Mishra}},
  \bibinfo {author} {\bibfnamefont {E.}~\bibnamefont {Avigad}}, \bibinfo
  {author} {\bibfnamefont {B.~P.}\ \bibnamefont {Bloom}}, \bibinfo {author}
  {\bibfnamefont {L.~T.}\ \bibnamefont {Baczewski}}, \bibinfo {author}
  {\bibfnamefont {S.}~\bibnamefont {Yochelis}}, \bibinfo {author}
  {\bibfnamefont {Y.}~\bibnamefont {Paltiel}}, \bibinfo {author} {\bibfnamefont
  {R.}~\bibnamefont {Naaman}}, \ and\ \bibinfo {author} {\bibfnamefont {D.~H.}\
  \bibnamefont {Waldeck}},\ }\bibfield  {title} {\enquote {\bibinfo {title}
  {Effect of chiral molecules on the electron's spin wavefunction at
  interfaces},}\ }\href {\doibase 10.1021/acs.jpclett.9b03487} {\bibfield
  {journal} {\bibinfo  {journal} {The Journal of Physical Chemistry Letters}\
  }\textbf {\bibinfo {volume} {11}},\ \bibinfo {pages} {1550--1557} (\bibinfo
  {year} {2020}{\natexlab{b}})}\BibitemShut {NoStop}%
\bibitem [{\citenamefont {Wolf}\ \emph {et~al.}(2022)\citenamefont {Wolf},
  \citenamefont {Liu}, \citenamefont {Xiao}, \citenamefont {Park},\ and\
  \citenamefont {Yan}}]{ACSNano.16.18601}%
  \BibitemOpen
  \bibfield  {author} {\bibinfo {author} {\bibfnamefont {Y.}~\bibnamefont
  {Wolf}}, \bibinfo {author} {\bibfnamefont {Y.}~\bibnamefont {Liu}}, \bibinfo
  {author} {\bibfnamefont {J.}~\bibnamefont {Xiao}}, \bibinfo {author}
  {\bibfnamefont {N.}~\bibnamefont {Park}}, \ and\ \bibinfo {author}
  {\bibfnamefont {B.}~\bibnamefont {Yan}},\ }\bibfield  {title} {\enquote
  {\bibinfo {title} {Unusual spin polarization in the chirality-induced spin
  selectivity},}\ }\href {\doibase 10.1021/acsnano.2c07088} {\bibfield
  {journal} {\bibinfo  {journal} {ACS Nano}\ }\textbf {\bibinfo {volume}
  {16}},\ \bibinfo {pages} {18601--18607} (\bibinfo {year} {2022})}\BibitemShut
  {NoStop}%
\bibitem [{\citenamefont {Adhikari}\ \emph {et~al.}(2023)\citenamefont
  {Adhikari}, \citenamefont {Liu}, \citenamefont {Wang}, \citenamefont {Hua},
  \citenamefont {Liu}, \citenamefont {Lochner}, \citenamefont {Schlottmann},
  \citenamefont {Yan}, \citenamefont {Zhao},\ and\ \citenamefont
  {Xiong}}]{NatComms.14.5163}%
  \BibitemOpen
  \bibfield  {author} {\bibinfo {author} {\bibfnamefont {Y.}~\bibnamefont
  {Adhikari}}, \bibinfo {author} {\bibfnamefont {T.}~\bibnamefont {Liu}},
  \bibinfo {author} {\bibfnamefont {H.}~\bibnamefont {Wang}}, \bibinfo {author}
  {\bibfnamefont {Z.}~\bibnamefont {Hua}}, \bibinfo {author} {\bibfnamefont
  {H.}~\bibnamefont {Liu}}, \bibinfo {author} {\bibfnamefont {E.}~\bibnamefont
  {Lochner}}, \bibinfo {author} {\bibfnamefont {P.}~\bibnamefont
  {Schlottmann}}, \bibinfo {author} {\bibfnamefont {B.}~\bibnamefont {Yan}},
  \bibinfo {author} {\bibfnamefont {J.}~\bibnamefont {Zhao}}, \ and\ \bibinfo
  {author} {\bibfnamefont {P.}~\bibnamefont {Xiong}},\ }\bibfield  {title}
  {\enquote {\bibinfo {title} {Interplay of structural chirality, electron spin
  and topological orbital in chiral molecular spin valves},}\ }\href {\doibase
  10.1038/s41467-023-40884-9} {\bibfield  {journal} {\bibinfo  {journal}
  {Nature Communications}\ }\textbf {\bibinfo {volume} {14}},\ \bibinfo {pages}
  {5163} (\bibinfo {year} {2023})}\BibitemShut {NoStop}%
\bibitem [{\citenamefont {Tirion}\ and\ \citenamefont {van
  Wees}(2024)}]{ACSNano.18.6028}%
  \BibitemOpen
  \bibfield  {author} {\bibinfo {author} {\bibfnamefont {S.~H.}\ \bibnamefont
  {Tirion}}\ and\ \bibinfo {author} {\bibfnamefont {B.~J.}\ \bibnamefont {van
  Wees}},\ }\bibfield  {title} {\enquote {\bibinfo {title} {Mechanism for
  electrostatically generated magnetoresistance in chiral systems without
  spin-dependent transport},}\ }\href {\doibase 10.1021/acsnano.3c12925}
  {\bibfield  {journal} {\bibinfo  {journal} {ACS Nano}\ }\textbf {\bibinfo
  {volume} {18}},\ \bibinfo {pages} {6028--6037} (\bibinfo {year}
  {2024})}\BibitemShut {NoStop}%
\bibitem [{\citenamefont {Zhao}\ \emph {et~al.}(2025)\citenamefont {Zhao},
  \citenamefont {Zhang}, \citenamefont {Xiao}, \citenamefont {Sun},\ and\
  \citenamefont {Yan}}]{NatComms.16.37}%
  \BibitemOpen
  \bibfield  {author} {\bibinfo {author} {\bibfnamefont {Y.}~\bibnamefont
  {Zhao}}, \bibinfo {author} {\bibfnamefont {K.}~\bibnamefont {Zhang}},
  \bibinfo {author} {\bibfnamefont {J.}~\bibnamefont {Xiao}}, \bibinfo {author}
  {\bibfnamefont {K.}~\bibnamefont {Sun}}, \ and\ \bibinfo {author}
  {\bibfnamefont {B.}~\bibnamefont {Yan}},\ }\bibfield  {title} {\enquote
  {\bibinfo {title} {Magnetochiral charge pumping due to charge trapping and
  skin effect in chirality-induced spin selectivity},}\ }\href {\doibase
  10.1038/s41467-024-55433-1} {\bibfield  {journal} {\bibinfo  {journal}
  {Nature Communications}\ }\textbf {\bibinfo {volume} {16}},\ \bibinfo {pages}
  {37} (\bibinfo {year} {2025})}\BibitemShut {NoStop}%
\bibitem [{\citenamefont {Dalum}\ and\ \citenamefont
  {Hedeg{\aa}rd}(2019)}]{NanoLett.19.5253}%
  \BibitemOpen
  \bibfield  {author} {\bibinfo {author} {\bibfnamefont {S.}~\bibnamefont
  {Dalum}}\ and\ \bibinfo {author} {\bibfnamefont {P.}~\bibnamefont
  {Hedeg{\aa}rd}},\ }\bibfield  {title} {\enquote {\bibinfo {title} {Theory of
  chiral induced spin selectivity},}\ }\href {\doibase
  10.1021/acs.nanolett.9b01707} {\bibfield  {journal} {\bibinfo  {journal}
  {Nano Letters}\ }\textbf {\bibinfo {volume} {19}},\ \bibinfo {pages}
  {5253--5259} (\bibinfo {year} {2019})}\BibitemShut {NoStop}%
\bibitem [{\citenamefont {Utsumi}, \citenamefont {Entin-Wohlman},\ and\
  \citenamefont {Aharony}(2020)}]{PhysRevB.102.035445}%
  \BibitemOpen
  \bibfield  {author} {\bibinfo {author} {\bibfnamefont {Y.}~\bibnamefont
  {Utsumi}}, \bibinfo {author} {\bibfnamefont {O.}~\bibnamefont
  {Entin-Wohlman}}, \ and\ \bibinfo {author} {\bibfnamefont {A.}~\bibnamefont
  {Aharony}},\ }\bibfield  {title} {\enquote {\bibinfo {title} {Spin
  selectivity through time-reversal symmetric helical junctions},}\ }\href
  {\doibase 10.1103/PhysRevB.102.035445} {\bibfield  {journal} {\bibinfo
  {journal} {Phys. Rev. B}\ }\textbf {\bibinfo {volume} {102}},\ \bibinfo
  {pages} {035445} (\bibinfo {year} {2020})}\BibitemShut {NoStop}%
\bibitem [{\citenamefont {Bardeen}, \citenamefont {Cooper},\ and\ \citenamefont
  {Schrieffer}(1957{\natexlab{a}})}]{PhysRev.106.162}%
  \BibitemOpen
  \bibfield  {author} {\bibinfo {author} {\bibfnamefont {J.}~\bibnamefont
  {Bardeen}}, \bibinfo {author} {\bibfnamefont {L.~N.}\ \bibnamefont {Cooper}},
  \ and\ \bibinfo {author} {\bibfnamefont {J.~R.}\ \bibnamefont {Schrieffer}},\
  }\bibfield  {title} {\enquote {\bibinfo {title} {Microscopic theory of
  superconductivity},}\ }\href {\doibase 10.1103/PhysRev.106.162} {\bibfield
  {journal} {\bibinfo  {journal} {Phys. Rev.}\ }\textbf {\bibinfo {volume}
  {106}},\ \bibinfo {pages} {162--164} (\bibinfo {year}
  {1957}{\natexlab{a}})}\BibitemShut {NoStop}%
\bibitem [{\citenamefont {Bardeen}, \citenamefont {Cooper},\ and\ \citenamefont
  {Schrieffer}(1957{\natexlab{b}})}]{PhysRev.108.1175}%
  \BibitemOpen
  \bibfield  {author} {\bibinfo {author} {\bibfnamefont {J.}~\bibnamefont
  {Bardeen}}, \bibinfo {author} {\bibfnamefont {L.~N.}\ \bibnamefont {Cooper}},
  \ and\ \bibinfo {author} {\bibfnamefont {J.~R.}\ \bibnamefont {Schrieffer}},\
  }\bibfield  {title} {\enquote {\bibinfo {title} {Theory of
  superconductivity},}\ }\href {\doibase 10.1103/PhysRev.108.1175} {\bibfield
  {journal} {\bibinfo  {journal} {Phys. Rev.}\ }\textbf {\bibinfo {volume}
  {108}},\ \bibinfo {pages} {1175--1204} (\bibinfo {year}
  {1957}{\natexlab{b}})}\BibitemShut {NoStop}%
\bibitem [{\citenamefont {M{\"o}llers}\ \emph {et~al.}(2022)\citenamefont
  {M{\"o}llers}, \citenamefont {Wei}, \citenamefont {Salamon}, \citenamefont
  {Bartsch}, \citenamefont {Wende}, \citenamefont {Waldeck},\ and\
  \citenamefont {Zacharias}}]{ACSNano.16.12145}%
  \BibitemOpen
  \bibfield  {author} {\bibinfo {author} {\bibfnamefont {P.~V.}\ \bibnamefont
  {M{\"o}llers}}, \bibinfo {author} {\bibfnamefont {J.}~\bibnamefont {Wei}},
  \bibinfo {author} {\bibfnamefont {S.}~\bibnamefont {Salamon}}, \bibinfo
  {author} {\bibfnamefont {M.}~\bibnamefont {Bartsch}}, \bibinfo {author}
  {\bibfnamefont {H.}~\bibnamefont {Wende}}, \bibinfo {author} {\bibfnamefont
  {D.~H.}\ \bibnamefont {Waldeck}}, \ and\ \bibinfo {author} {\bibfnamefont
  {H.}~\bibnamefont {Zacharias}},\ }\bibfield  {title} {\enquote {\bibinfo
  {title} {Spin-polarized photoemission from chiral cuo catalyst thin films},}\
  }\href {\doibase 10.1021/acsnano.2c02709} {\bibfield  {journal} {\bibinfo
  {journal} {ACS Nano}\ }\textbf {\bibinfo {volume} {16}},\ \bibinfo {pages}
  {12145--12155} (\bibinfo {year} {2022})}\BibitemShut {NoStop}%
\bibitem [{\citenamefont {Xie}\ \emph {et~al.}(2011)\citenamefont {Xie},
  \citenamefont {Markus}, \citenamefont {Cohen}, \citenamefont {Vager},
  \citenamefont {Gutierrez},\ and\ \citenamefont {Naaman}}]{NanoLett.11.4652}%
  \BibitemOpen
  \bibfield  {author} {\bibinfo {author} {\bibfnamefont {Z.}~\bibnamefont
  {Xie}}, \bibinfo {author} {\bibfnamefont {T.~Z.}\ \bibnamefont {Markus}},
  \bibinfo {author} {\bibfnamefont {S.~R.}\ \bibnamefont {Cohen}}, \bibinfo
  {author} {\bibfnamefont {Z.}~\bibnamefont {Vager}}, \bibinfo {author}
  {\bibfnamefont {R.}~\bibnamefont {Gutierrez}}, \ and\ \bibinfo {author}
  {\bibfnamefont {R.}~\bibnamefont {Naaman}},\ }\bibfield  {title} {\enquote
  {\bibinfo {title} {Spin specific electron conduction through dna
  oligomers},}\ }\href {\doibase 10.1021/nl2021637} {\bibfield  {journal}
  {\bibinfo  {journal} {Nano Letters}\ }\textbf {\bibinfo {volume} {11}},\
  \bibinfo {pages} {4652--4655} (\bibinfo {year} {2011})}\BibitemShut {NoStop}%
\bibitem [{\citenamefont {Kiran}\ \emph {et~al.}(2016)\citenamefont {Kiran},
  \citenamefont {Mathew}, \citenamefont {Cohen}, \citenamefont
  {Hern{\'a}ndez~Delgado}, \citenamefont {Lacour},\ and\ \citenamefont
  {Naaman}}]{AdvMater.28.1957}%
  \BibitemOpen
  \bibfield  {author} {\bibinfo {author} {\bibfnamefont {V.}~\bibnamefont
  {Kiran}}, \bibinfo {author} {\bibfnamefont {S.~P.}\ \bibnamefont {Mathew}},
  \bibinfo {author} {\bibfnamefont {S.~R.}\ \bibnamefont {Cohen}}, \bibinfo
  {author} {\bibfnamefont {I.}~\bibnamefont {Hern{\'a}ndez~Delgado}}, \bibinfo
  {author} {\bibfnamefont {J.}~\bibnamefont {Lacour}}, \ and\ \bibinfo {author}
  {\bibfnamefont {R.}~\bibnamefont {Naaman}},\ }\bibfield  {title} {\enquote
  {\bibinfo {title} {Helicenes---a new class of organic spin filter},}\ }\href
  {\doibase https://doi.org/10.1002/adma.201504725} {\bibfield  {journal}
  {\bibinfo  {journal} {Advanced Materials}\ }\textbf {\bibinfo {volume}
  {28}},\ \bibinfo {pages} {1957--1962} (\bibinfo {year} {2016})}\BibitemShut
  {NoStop}%
\bibitem [{\citenamefont {Safari}\ \emph {et~al.}(2024)\citenamefont {Safari},
  \citenamefont {Matthes}, \citenamefont {Schneider}, \citenamefont {Ernst},\
  and\ \citenamefont {B{\"u}rgler}}]{Small.20.2308233}%
  \BibitemOpen
  \bibfield  {author} {\bibinfo {author} {\bibfnamefont {M.~R.}\ \bibnamefont
  {Safari}}, \bibinfo {author} {\bibfnamefont {F.}~\bibnamefont {Matthes}},
  \bibinfo {author} {\bibfnamefont {C.~M.}\ \bibnamefont {Schneider}}, \bibinfo
  {author} {\bibfnamefont {K.-H.}\ \bibnamefont {Ernst}}, \ and\ \bibinfo
  {author} {\bibfnamefont {D.~E.}\ \bibnamefont {B{\"u}rgler}},\ }\bibfield
  {title} {\enquote {\bibinfo {title} {Spin-selective electron transport
  through single chiral molecules},}\ }\href {\doibase
  https://doi.org/10.1002/smll.202308233} {\bibfield  {journal} {\bibinfo
  {journal} {Small}\ }\textbf {\bibinfo {volume} {20}},\ \bibinfo {pages}
  {2308233} (\bibinfo {year} {2024})}\BibitemShut {NoStop}%
\bibitem [{\citenamefont {Onsager}(1931)}]{PhysRev.37.405}%
  \BibitemOpen
  \bibfield  {author} {\bibinfo {author} {\bibfnamefont {L.}~\bibnamefont
  {Onsager}},\ }\bibfield  {title} {\enquote {\bibinfo {title} {Reciprocal
  relations in irreversible processes. i.}}\ }\href {\doibase
  10.1103/PhysRev.37.405} {\bibfield  {journal} {\bibinfo  {journal} {Phys.
  Rev.}\ }\textbf {\bibinfo {volume} {37}},\ \bibinfo {pages} {405--426}
  (\bibinfo {year} {1931})}\BibitemShut {NoStop}%
\bibitem [{\citenamefont {Fransson}(2020)}]{PhysRevB.102.235416}%
  \BibitemOpen
  \bibfield  {author} {\bibinfo {author} {\bibfnamefont {J.}~\bibnamefont
  {Fransson}},\ }\bibfield  {title} {\enquote {\bibinfo {title} {Vibrational
  origin of exchange splitting and chiral-induced spin selectivity},}\ }\href
  {\doibase 10.1103/PhysRevB.102.235416} {\bibfield  {journal} {\bibinfo
  {journal} {Phys. Rev. B}\ }\textbf {\bibinfo {volume} {102}},\ \bibinfo
  {pages} {235416} (\bibinfo {year} {2020})}\BibitemShut {NoStop}%
\bibitem [{\citenamefont {Liu}\ \emph {et~al.}(2021)\citenamefont {Liu},
  \citenamefont {Xiao}, \citenamefont {Koo},\ and\ \citenamefont
  {Yan}}]{NatMaterials.5.638}%
  \BibitemOpen
  \bibfield  {author} {\bibinfo {author} {\bibfnamefont {Y.}~\bibnamefont
  {Liu}}, \bibinfo {author} {\bibfnamefont {J.}~\bibnamefont {Xiao}}, \bibinfo
  {author} {\bibfnamefont {J.}~\bibnamefont {Koo}}, \ and\ \bibinfo {author}
  {\bibfnamefont {B.}~\bibnamefont {Yan}},\ }\bibfield  {title} {\enquote
  {\bibinfo {title} {Chirality-driven topological electronic structure of
  dna-like materials},}\ }\href {\doibase 10.1038/s41563-021-00924-5}
  {\bibfield  {journal} {\bibinfo  {journal} {Nature Materials}\ }\textbf
  {\bibinfo {volume} {20}},\ \bibinfo {pages} {638--644} (\bibinfo {year}
  {2021})}\BibitemShut {NoStop}%
\bibitem [{\citenamefont {Smolinsky}\ \emph {et~al.}(2019)\citenamefont
  {Smolinsky}, \citenamefont {Neubauer}, \citenamefont {Kumar}, \citenamefont
  {Yochelis}, \citenamefont {Capua}, \citenamefont {Carmieli}, \citenamefont
  {Paltiel}, \citenamefont {Naaman},\ and\ \citenamefont
  {Michaeli}}]{JPhysChemLett.10.1139}%
  \BibitemOpen
  \bibfield  {author} {\bibinfo {author} {\bibfnamefont {E.~Z.~B.}\
  \bibnamefont {Smolinsky}}, \bibinfo {author} {\bibfnamefont {A.}~\bibnamefont
  {Neubauer}}, \bibinfo {author} {\bibfnamefont {A.}~\bibnamefont {Kumar}},
  \bibinfo {author} {\bibfnamefont {S.}~\bibnamefont {Yochelis}}, \bibinfo
  {author} {\bibfnamefont {E.}~\bibnamefont {Capua}}, \bibinfo {author}
  {\bibfnamefont {R.}~\bibnamefont {Carmieli}}, \bibinfo {author}
  {\bibfnamefont {Y.}~\bibnamefont {Paltiel}}, \bibinfo {author} {\bibfnamefont
  {R.}~\bibnamefont {Naaman}}, \ and\ \bibinfo {author} {\bibfnamefont
  {K.}~\bibnamefont {Michaeli}},\ }\bibfield  {title} {\enquote {\bibinfo
  {title} {Electric field-controlled magnetization in gaas/algaas
  heterostructures--chiral organic molecules hybrids},}\ }\href {\doibase
  10.1021/acs.jpclett.9b00092} {\bibfield  {journal} {\bibinfo  {journal} {The
  Journal of Physical Chemistry Letters}\ }\textbf {\bibinfo {volume} {10}},\
  \bibinfo {pages} {1139--1145} (\bibinfo {year} {2019})},\ \Eprint
  {http://arxiv.org/abs/https://doi.org/10.1021/acs.jpclett.9b00092}
  {https://doi.org/10.1021/acs.jpclett.9b00092} \BibitemShut {NoStop}%
\bibitem [{\citenamefont {Fridman}\ \emph {et~al.}(2019)\citenamefont
  {Fridman}, \citenamefont {Dehnel}, \citenamefont {Yochelis}, \citenamefont
  {Lifshitz},\ and\ \citenamefont {Paltiel}}]{JPhysChemLett.10.3858}%
  \BibitemOpen
  \bibfield  {author} {\bibinfo {author} {\bibfnamefont {H.~T.}\ \bibnamefont
  {Fridman}}, \bibinfo {author} {\bibfnamefont {J.}~\bibnamefont {Dehnel}},
  \bibinfo {author} {\bibfnamefont {S.}~\bibnamefont {Yochelis}}, \bibinfo
  {author} {\bibfnamefont {E.}~\bibnamefont {Lifshitz}}, \ and\ \bibinfo
  {author} {\bibfnamefont {Y.}~\bibnamefont {Paltiel}},\ }\bibfield  {title}
  {\enquote {\bibinfo {title} {Spin-exciton delocalization enhancement in
  multilayer chiral linker/quantum dot structures},}\ }\href {\doibase
  10.1021/acs.jpclett.9b01433} {\bibfield  {journal} {\bibinfo  {journal} {The
  Journal of Physical Chemistry Letters}\ }\textbf {\bibinfo {volume} {10}},\
  \bibinfo {pages} {3858--3862} (\bibinfo {year} {2019})}\BibitemShut {NoStop}%
\bibitem [{\citenamefont {Zhu}\ \emph {et~al.}(2024)\citenamefont {Zhu},
  \citenamefont {Cohen}, \citenamefont {Brontvein}, \citenamefont {Fransson},\
  and\ \citenamefont {Naaman}}]{Small.20.2406631}%
  \BibitemOpen
  \bibfield  {author} {\bibinfo {author} {\bibfnamefont {Q.}~\bibnamefont
  {Zhu}}, \bibinfo {author} {\bibfnamefont {S.~R.}\ \bibnamefont {Cohen}},
  \bibinfo {author} {\bibfnamefont {O.}~\bibnamefont {Brontvein}}, \bibinfo
  {author} {\bibfnamefont {J.}~\bibnamefont {Fransson}}, \ and\ \bibinfo
  {author} {\bibfnamefont {R.}~\bibnamefont {Naaman}},\ }\bibfield  {title}
  {\enquote {\bibinfo {title} {Magnetic monopole-like behavior in
  superparamagnetic nanoparticle coated with chiral molecules},}\ }\href
  {\doibase https://doi.org/10.1002/smll.202406631} {\bibfield  {journal}
  {\bibinfo  {journal} {Small}\ }\textbf {\bibinfo {volume} {20}},\ \bibinfo
  {pages} {2406631} (\bibinfo {year} {2024})}\BibitemShut {NoStop}%
\bibitem [{\citenamefont {Kubo}(1957)}]{JPhysSocJpn.12.570}%
  \BibitemOpen
  \bibfield  {author} {\bibinfo {author} {\bibfnamefont {R.}~\bibnamefont
  {Kubo}},\ }\bibfield  {title} {\enquote {\bibinfo {title}
  {Statistical-mechanical theory of irreversible processes. i. general theory
  and simple applications to magnetic and conduction problems},}\ }\href
  {\doibase 10.1143/JPSJ.12.570} {\bibfield  {journal} {\bibinfo  {journal}
  {Journal of the Physical Society of Japan}\ }\textbf {\bibinfo {volume}
  {12}},\ \bibinfo {pages} {570--586} (\bibinfo {year} {1957})}\BibitemShut
  {NoStop}%
\bibitem [{\citenamefont {Fransson}(2019)}]{JPhysChemLett.10.7126}%
  \BibitemOpen
  \bibfield  {author} {\bibinfo {author} {\bibfnamefont {J.}~\bibnamefont
  {Fransson}},\ }\bibfield  {title} {\enquote {\bibinfo {title}
  {Chirality-induced spin selectivity: The role of electron correlations},}\
  }\href {\doibase 10.1021/acs.jpclett.9b02929} {\bibfield  {journal} {\bibinfo
   {journal} {The Journal of Physical Chemistry Letters}\ }\textbf {\bibinfo
  {volume} {10}},\ \bibinfo {pages} {7126--7132} (\bibinfo {year}
  {2019})}\BibitemShut {NoStop}%
\bibitem [{\citenamefont {Zhang}\ \emph {et~al.}(2020)\citenamefont {Zhang},
  \citenamefont {Hao}, \citenamefont {Qin}, \citenamefont {Xie},\ and\
  \citenamefont {Qu}}]{PhysRevB.102.214303}%
  \BibitemOpen
  \bibfield  {author} {\bibinfo {author} {\bibfnamefont {L.}~\bibnamefont
  {Zhang}}, \bibinfo {author} {\bibfnamefont {Y.}~\bibnamefont {Hao}}, \bibinfo
  {author} {\bibfnamefont {W.}~\bibnamefont {Qin}}, \bibinfo {author}
  {\bibfnamefont {S.}~\bibnamefont {Xie}}, \ and\ \bibinfo {author}
  {\bibfnamefont {F.}~\bibnamefont {Qu}},\ }\bibfield  {title} {\enquote
  {\bibinfo {title} {Chiral-induced spin selectivity: A polaron transport
  model},}\ }\href {\doibase 10.1103/PhysRevB.102.214303} {\bibfield  {journal}
  {\bibinfo  {journal} {Phys. Rev. B}\ }\textbf {\bibinfo {volume} {102}},\
  \bibinfo {pages} {214303} (\bibinfo {year} {2020})}\BibitemShut {NoStop}%
\bibitem [{\citenamefont {Alwan}\ and\ \citenamefont
  {Dubi}(2021)}]{JACS.143.14235}%
  \BibitemOpen
  \bibfield  {author} {\bibinfo {author} {\bibfnamefont {S.}~\bibnamefont
  {Alwan}}\ and\ \bibinfo {author} {\bibfnamefont {Y.}~\bibnamefont {Dubi}},\
  }\bibfield  {title} {\enquote {\bibinfo {title} {Spinterface origin for the
  chirality-induced spin-selectivity effect},}\ }\href {\doibase
  10.1021/jacs.1c05637} {\bibfield  {journal} {\bibinfo  {journal} {Journal of
  the American Chemical Society}\ }\textbf {\bibinfo {volume} {143}},\ \bibinfo
  {pages} {14235--14241} (\bibinfo {year} {2021})}\BibitemShut {NoStop}%
\bibitem [{\citenamefont {Huisman}\ and\ \citenamefont
  {Thijssen}(2021)}]{JPhysChemC.125.23364}%
  \BibitemOpen
  \bibfield  {author} {\bibinfo {author} {\bibfnamefont {K.~H.}\ \bibnamefont
  {Huisman}}\ and\ \bibinfo {author} {\bibfnamefont {J.~M.}\ \bibnamefont
  {Thijssen}},\ }\bibfield  {title} {\enquote {\bibinfo {title} {Ciss effect: A
  magnetoresistance through inelastic scattering},}\ }\href {\doibase
  10.1021/acs.jpcc.1c06193} {\bibfield  {journal} {\bibinfo  {journal} {The
  Journal of Physical Chemistry C}\ }\textbf {\bibinfo {volume} {125}},\
  \bibinfo {pages} {23364--23369} (\bibinfo {year} {2021})}\BibitemShut
  {NoStop}%
\bibitem [{\citenamefont {Huisman}, \citenamefont {Heinisch},\ and\
  \citenamefont {Thijssen}(2023)}]{JPhysChemC.127.6900}%
  \BibitemOpen
  \bibfield  {author} {\bibinfo {author} {\bibfnamefont {K.~H.}\ \bibnamefont
  {Huisman}}, \bibinfo {author} {\bibfnamefont {J.-B. M.-Y.}\ \bibnamefont
  {Heinisch}}, \ and\ \bibinfo {author} {\bibfnamefont {J.~M.}\ \bibnamefont
  {Thijssen}},\ }\bibfield  {title} {\enquote {\bibinfo {title}
  {Chirality-induced spin selectivity (ciss) effect: Magnetocurrent--voltage
  characteristics with coulomb interactions i},}\ }\href {\doibase
  10.1021/acs.jpcc.2c08807} {\bibfield  {journal} {\bibinfo  {journal} {The
  Journal of Physical Chemistry C}\ }\textbf {\bibinfo {volume} {127}},\
  \bibinfo {pages} {6900--6905} (\bibinfo {year} {2023})}\BibitemShut {NoStop}%
\bibitem [{\citenamefont {Vittmann}\ \emph {et~al.}(2023)\citenamefont
  {Vittmann}, \citenamefont {Lim}, \citenamefont {Tamascelli}, \citenamefont
  {Huelga},\ and\ \citenamefont {Plenio}}]{JPhysChemLett.14.340}%
  \BibitemOpen
  \bibfield  {author} {\bibinfo {author} {\bibfnamefont {C.}~\bibnamefont
  {Vittmann}}, \bibinfo {author} {\bibfnamefont {J.}~\bibnamefont {Lim}},
  \bibinfo {author} {\bibfnamefont {D.}~\bibnamefont {Tamascelli}}, \bibinfo
  {author} {\bibfnamefont {S.~F.}\ \bibnamefont {Huelga}}, \ and\ \bibinfo
  {author} {\bibfnamefont {M.~B.}\ \bibnamefont {Plenio}},\ }\bibfield  {title}
  {\enquote {\bibinfo {title} {Spin-dependent momentum conservation of
  electron--phonon scattering in chirality-induced spin selectivity},}\ }\href
  {\doibase 10.1021/acs.jpclett.2c03224} {\bibfield  {journal} {\bibinfo
  {journal} {The Journal of Physical Chemistry Letters}\ }\textbf {\bibinfo
  {volume} {14}},\ \bibinfo {pages} {340--346} (\bibinfo {year} {2023})},\
  \bibinfo {note} {pMID: 36625481}\BibitemShut {NoStop}%
\bibitem [{\citenamefont {Alwan}, \citenamefont {Sharoni},\ and\ \citenamefont
  {Dubi}(2024)}]{JPhysChemC.128.6438}%
  \BibitemOpen
  \bibfield  {author} {\bibinfo {author} {\bibfnamefont {S.}~\bibnamefont
  {Alwan}}, \bibinfo {author} {\bibfnamefont {A.}~\bibnamefont {Sharoni}}, \
  and\ \bibinfo {author} {\bibfnamefont {Y.}~\bibnamefont {Dubi}},\ }\bibfield
  {title} {\enquote {\bibinfo {title} {Role of electrode polarization in the
  electron transport chirality-induced spin-selectivity effect},}\ }\href
  {\doibase 10.1021/acs.jpcc.3c08223} {\bibfield  {journal} {\bibinfo
  {journal} {The Journal of Physical Chemistry C}\ }\textbf {\bibinfo {volume}
  {128}},\ \bibinfo {pages} {6438--6445} (\bibinfo {year} {2024})}\BibitemShut
  {NoStop}%
\bibitem [{\citenamefont
  {Fransson}(2025{\natexlab{a}})}]{JPhysChemLett.16.4346}%
  \BibitemOpen
  \bibfield  {author} {\bibinfo {author} {\bibfnamefont {J.}~\bibnamefont
  {Fransson}},\ }\bibfield  {title} {\enquote {\bibinfo {title} {Chiral induced
  spin polarized electron current: Origin of the chiral induced spin
  selectivity effect},}\ }\href {\doibase 10.1021/acs.jpclett.5c00104}
  {\bibfield  {journal} {\bibinfo  {journal} {The Journal of Physical Chemistry
  Letters}\ }\textbf {\bibinfo {volume} {16}},\ \bibinfo {pages} {4346--4353}
  (\bibinfo {year} {2025}{\natexlab{a}})}\BibitemShut {NoStop}%
\bibitem [{\citenamefont {Savi}\ \emph {et~al.}(2025)\citenamefont {Savi},
  \citenamefont {Celada}, \citenamefont {Phan~Huu}, \citenamefont {Chiesa},
  \citenamefont {Carretta},\ and\ \citenamefont
  {Painelli}}]{JPhysChemLett.16.9107}%
  \BibitemOpen
  \bibfield  {author} {\bibinfo {author} {\bibfnamefont {L.}~\bibnamefont
  {Savi}}, \bibinfo {author} {\bibfnamefont {L.}~\bibnamefont {Celada}},
  \bibinfo {author} {\bibfnamefont {D.~K.~A.}\ \bibnamefont {Phan~Huu}},
  \bibinfo {author} {\bibfnamefont {A.}~\bibnamefont {Chiesa}}, \bibinfo
  {author} {\bibfnamefont {S.}~\bibnamefont {Carretta}}, \ and\ \bibinfo
  {author} {\bibfnamefont {A.}~\bibnamefont {Painelli}},\ }\bibfield  {title}
  {\enquote {\bibinfo {title} {Chirality-induced spin selectivity: A minimal
  model},}\ }\href {\doibase 10.1021/acs.jpclett.5c01813} {\bibfield  {journal}
  {\bibinfo  {journal} {The Journal of Physical Chemistry Letters}\ }\textbf
  {\bibinfo {volume} {16}},\ \bibinfo {pages} {9107--9115} (\bibinfo {year}
  {2025})}\BibitemShut {NoStop}%
\bibitem [{\citenamefont {Fransson}(2025{\natexlab{b}})}]{arXiv:2509.17817}%
  \BibitemOpen
  \bibfield  {author} {\bibinfo {author} {\bibfnamefont {J.}~\bibnamefont
  {Fransson}},\ }\href {https://arxiv.org/abs/2509.17817} {\enquote {\bibinfo
  {title} {Breaking of time-reversal symmetry and onsager reciprocity in chiral
  molecule interfacd with an environment},}\ } (\bibinfo {year}
  {2025}{\natexlab{b}}),\ \Eprint {http://arxiv.org/abs/2509.17817}
  {arXiv:2509.17817 [cond-mat.mes-hall]} \BibitemShut {NoStop}%
\bibitem [{\citenamefont {Renaud}\ \emph {et~al.}(2013)\citenamefont {Renaud},
  \citenamefont {Berlin}, \citenamefont {Lewis},\ and\ \citenamefont
  {Ratner}}]{JACS.135.3953}%
  \BibitemOpen
  \bibfield  {author} {\bibinfo {author} {\bibfnamefont {N.}~\bibnamefont
  {Renaud}}, \bibinfo {author} {\bibfnamefont {Y.~A.}\ \bibnamefont {Berlin}},
  \bibinfo {author} {\bibfnamefont {F.~D.}\ \bibnamefont {Lewis}}, \ and\
  \bibinfo {author} {\bibfnamefont {M.~A.}\ \bibnamefont {Ratner}},\ }\bibfield
   {title} {\enquote {\bibinfo {title} {Between superexchange and hopping: An
  intermediate charge-transfer mechanism in poly(a)-poly(t) dna hairpins},}\
  }\href {\doibase 10.1021/ja3113998} {\bibfield  {journal} {\bibinfo
  {journal} {Journal of the American Chemical Society}\ }\textbf {\bibinfo
  {volume} {135}},\ \bibinfo {pages} {3953--3963} (\bibinfo {year}
  {2013})}\BibitemShut {NoStop}%
\bibitem [{\citenamefont {Li}\ \emph {et~al.}(2015)\citenamefont {Li},
  \citenamefont {Govind}, \citenamefont {Ratner}, \citenamefont {Cramer},\ and\
  \citenamefont {Gagliardi}}]{JPhysChemLett.6.4889}%
  \BibitemOpen
  \bibfield  {author} {\bibinfo {author} {\bibfnamefont {G.}~\bibnamefont
  {Li}}, \bibinfo {author} {\bibfnamefont {N.}~\bibnamefont {Govind}}, \bibinfo
  {author} {\bibfnamefont {M.~A.}\ \bibnamefont {Ratner}}, \bibinfo {author}
  {\bibfnamefont {C.~J.}\ \bibnamefont {Cramer}}, \ and\ \bibinfo {author}
  {\bibfnamefont {L.}~\bibnamefont {Gagliardi}},\ }\bibfield  {title} {\enquote
  {\bibinfo {title} {Influence of coherent tunneling and incoherent hopping on
  the charge transfer mechanism in linear donor--bridge--acceptor systems},}\
  }\href {\doibase 10.1021/acs.jpclett.5b02154} {\bibfield  {journal} {\bibinfo
   {journal} {The Journal of Physical Chemistry Letters}\ }\textbf {\bibinfo
  {volume} {6}},\ \bibinfo {pages} {4889--4897} (\bibinfo {year}
  {2015})}\BibitemShut {NoStop}%
\bibitem [{\citenamefont {Geyer}, \citenamefont {Gutierrez},\ and\
  \citenamefont {Cuniberti}(2020)}]{JChemPhys.152.214105}%
  \BibitemOpen
  \bibfield  {author} {\bibinfo {author} {\bibfnamefont {M.}~\bibnamefont
  {Geyer}}, \bibinfo {author} {\bibfnamefont {R.}~\bibnamefont {Gutierrez}}, \
  and\ \bibinfo {author} {\bibfnamefont {G.}~\bibnamefont {Cuniberti}},\
  }\bibfield  {title} {\enquote {\bibinfo {title} {Effective hamiltonian model
  for helically constrained quantum systems within adiabatic perturbation
  theory: Application to the chirality-induced spin selectivity (ciss)
  effect},}\ }\href {\doibase 10.1063/5.0005181} {\bibfield  {journal}
  {\bibinfo  {journal} {The Journal of Chemical Physics}\ }\textbf {\bibinfo
  {volume} {152}},\ \bibinfo {pages} {214105} (\bibinfo {year} {2020})},\
  \Eprint
  {http://arxiv.org/abs/https://pubs.aip.org/aip/jcp/article-pdf/doi/10.1063/5.0005181/15575532/214105\_1\_online.pdf}
  {https://pubs.aip.org/aip/jcp/article-pdf/doi/10.1063/5.0005181/15575532/214105\_1\_online.pdf}
  \BibitemShut {NoStop}%
\bibitem [{\citenamefont {Mishra}\ \emph {et~al.}(2020)\citenamefont {Mishra},
  \citenamefont {Mondal}, \citenamefont {Pal}, \citenamefont {Das},
  \citenamefont {Smolinsky}, \citenamefont {Siligardi},\ and\ \citenamefont
  {Naaman}}]{JPhysChemC.124.10776}%
  \BibitemOpen
  \bibfield  {author} {\bibinfo {author} {\bibfnamefont {S.}~\bibnamefont
  {Mishra}}, \bibinfo {author} {\bibfnamefont {A.~K.}\ \bibnamefont {Mondal}},
  \bibinfo {author} {\bibfnamefont {S.}~\bibnamefont {Pal}}, \bibinfo {author}
  {\bibfnamefont {T.~K.}\ \bibnamefont {Das}}, \bibinfo {author} {\bibfnamefont
  {E.~Z.~B.}\ \bibnamefont {Smolinsky}}, \bibinfo {author} {\bibfnamefont
  {G.}~\bibnamefont {Siligardi}}, \ and\ \bibinfo {author} {\bibfnamefont
  {R.}~\bibnamefont {Naaman}},\ }\bibfield  {title} {\enquote {\bibinfo {title}
  {Length-dependent electron spin polarization in oligopeptides and dna},}\
  }\href {\doibase 10.1021/acs.jpcc.0c02291} {\bibfield  {journal} {\bibinfo
  {journal} {The Journal of Physical Chemistry C}\ }\textbf {\bibinfo {volume}
  {124}},\ \bibinfo {pages} {10776--10782} (\bibinfo {year}
  {2020})}\BibitemShut {NoStop}%
\bibitem [{\citenamefont {Hossain}\ \emph {et~al.}(2023)\citenamefont
  {Hossain}, \citenamefont {Illescas-Lopez}, \citenamefont {Nair},
  \citenamefont {Cuerva}, \citenamefont {Álvarez~de Cienfuegos},\ and\
  \citenamefont {Pramanik}}]{NanoscaleHoriz.8.320}%
  \BibitemOpen
  \bibfield  {author} {\bibinfo {author} {\bibfnamefont {M.~A.}\ \bibnamefont
  {Hossain}}, \bibinfo {author} {\bibfnamefont {S.}~\bibnamefont
  {Illescas-Lopez}}, \bibinfo {author} {\bibfnamefont {R.}~\bibnamefont
  {Nair}}, \bibinfo {author} {\bibfnamefont {J.~M.}\ \bibnamefont {Cuerva}},
  \bibinfo {author} {\bibfnamefont {L.}~\bibnamefont {Álvarez~de Cienfuegos}},
  \ and\ \bibinfo {author} {\bibfnamefont {S.}~\bibnamefont {Pramanik}},\
  }\bibfield  {title} {\enquote {\bibinfo {title} {Transverse
  magnetoconductance in two-terminal chiral spin-selective devices},}\ }\href
  {\doibase 10.1039/D2NH00502F} {\bibfield  {journal} {\bibinfo  {journal}
  {Nanoscale Horiz.}\ }\textbf {\bibinfo {volume} {8}},\ \bibinfo {pages}
  {320--330} (\bibinfo {year} {2023})}\BibitemShut {NoStop}%
\bibitem [{\citenamefont {Singh~Bisht}\ \emph {et~al.}(2024)\citenamefont
  {Singh~Bisht}, \citenamefont {Garg}, \citenamefont {Nakka},\ and\
  \citenamefont {Mondal}}]{JPhysChemLett.15.6605}%
  \BibitemOpen
  \bibfield  {author} {\bibinfo {author} {\bibfnamefont {P.}~\bibnamefont
  {Singh~Bisht}}, \bibinfo {author} {\bibfnamefont {R.}~\bibnamefont {Garg}},
  \bibinfo {author} {\bibfnamefont {N.}~\bibnamefont {Nakka}}, \ and\ \bibinfo
  {author} {\bibfnamefont {A.~K.}\ \bibnamefont {Mondal}},\ }\bibfield  {title}
  {\enquote {\bibinfo {title} {Spin filtering and amplification in
  self-assembled nanofibers based on chiral asymmetric building blocks},}\
  }\href {\doibase 10.1021/acs.jpclett.4c01423} {\bibfield  {journal} {\bibinfo
   {journal} {The Journal of Physical Chemistry Letters}\ }\textbf {\bibinfo
  {volume} {15}},\ \bibinfo {pages} {6605--6610} (\bibinfo {year}
  {2024})}\BibitemShut {NoStop}%
\bibitem [{\citenamefont {Albro}\ \emph {et~al.}(2025)\citenamefont {Albro},
  \citenamefont {Garrett}, \citenamefont {Govindaraj}, \citenamefont {Bloom},
  \citenamefont {Rosi},\ and\ \citenamefont {Waldeck}}]{ACSNano.19.17941}%
  \BibitemOpen
  \bibfield  {author} {\bibinfo {author} {\bibfnamefont {J.~A.}\ \bibnamefont
  {Albro}}, \bibinfo {author} {\bibfnamefont {N.~T.}\ \bibnamefont {Garrett}},
  \bibinfo {author} {\bibfnamefont {K.}~\bibnamefont {Govindaraj}}, \bibinfo
  {author} {\bibfnamefont {B.~P.}\ \bibnamefont {Bloom}}, \bibinfo {author}
  {\bibfnamefont {N.~L.}\ \bibnamefont {Rosi}}, \ and\ \bibinfo {author}
  {\bibfnamefont {D.~H.}\ \bibnamefont {Waldeck}},\ }\bibfield  {title}
  {\enquote {\bibinfo {title} {Measurement platform to probe the mechanism of
  chiral-induced spin selectivity through direction-dependent magnetic
  conductive atomic force microscopy},}\ }\href {\doibase
  10.1021/acsnano.5c04980} {\bibfield  {journal} {\bibinfo  {journal} {ACS
  Nano}\ }\textbf {\bibinfo {volume} {19}},\ \bibinfo {pages} {17941--17949}
  (\bibinfo {year} {2025})}\BibitemShut {NoStop}%
\bibitem [{\citenamefont {Fransson}\ \emph {et~al.}(2017)\citenamefont
  {Fransson}, \citenamefont {Thonig}, \citenamefont {Bessarab}, \citenamefont
  {Bhattacharjee}, \citenamefont {Hellsvik},\ and\ \citenamefont
  {Nordstr\"om}}]{PhysRevMaterials.1.074404}%
  \BibitemOpen
  \bibfield  {author} {\bibinfo {author} {\bibfnamefont {J.}~\bibnamefont
  {Fransson}}, \bibinfo {author} {\bibfnamefont {D.}~\bibnamefont {Thonig}},
  \bibinfo {author} {\bibfnamefont {P.~F.}\ \bibnamefont {Bessarab}}, \bibinfo
  {author} {\bibfnamefont {S.}~\bibnamefont {Bhattacharjee}}, \bibinfo {author}
  {\bibfnamefont {J.}~\bibnamefont {Hellsvik}}, \ and\ \bibinfo {author}
  {\bibfnamefont {L.}~\bibnamefont {Nordstr\"om}},\ }\bibfield  {title}
  {\enquote {\bibinfo {title} {Microscopic theory for coupled atomistic
  magnetization and lattice dynamics},}\ }\href {\doibase
  10.1103/PhysRevMaterials.1.074404} {\bibfield  {journal} {\bibinfo  {journal}
  {Phys. Rev. Materials}\ }\textbf {\bibinfo {volume} {1}},\ \bibinfo {pages}
  {074404} (\bibinfo {year} {2017})}\BibitemShut {NoStop}%
\bibitem [{\citenamefont {Fransson}(2023)}]{PhysRevResearch.5.L022039}%
  \BibitemOpen
  \bibfield  {author} {\bibinfo {author} {\bibfnamefont {J.}~\bibnamefont
  {Fransson}},\ }\bibfield  {title} {\enquote {\bibinfo {title} {Chiral phonon
  induced spin polarization},}\ }\href {\doibase
  10.1103/PhysRevResearch.5.L022039} {\bibfield  {journal} {\bibinfo  {journal}
  {Phys. Rev. Res.}\ }\textbf {\bibinfo {volume} {5}},\ \bibinfo {pages}
  {L022039} (\bibinfo {year} {2023})}\BibitemShut {NoStop}%
\bibitem [{\citenamefont {Dou}, \citenamefont {Miao},\ and\ \citenamefont
  {Subotnik}(2017)}]{PhysRevLett.119.046001}%
  \BibitemOpen
  \bibfield  {author} {\bibinfo {author} {\bibfnamefont {W.}~\bibnamefont
  {Dou}}, \bibinfo {author} {\bibfnamefont {G.}~\bibnamefont {Miao}}, \ and\
  \bibinfo {author} {\bibfnamefont {J.~E.}\ \bibnamefont {Subotnik}},\
  }\bibfield  {title} {\enquote {\bibinfo {title} {Born-oppenheimer dynamics,
  electronic friction, and the inclusion of electron-electron interactions},}\
  }\href {\doibase 10.1103/PhysRevLett.119.046001} {\bibfield  {journal}
  {\bibinfo  {journal} {Phys. Rev. Lett.}\ }\textbf {\bibinfo {volume} {119}},\
  \bibinfo {pages} {046001} (\bibinfo {year} {2017})}\BibitemShut {NoStop}%
\bibitem [{\citenamefont {Teh}, \citenamefont {Dou},\ and\ \citenamefont
  {Subotnik}(2021)}]{PhysRevB.104.L201409}%
  \BibitemOpen
  \bibfield  {author} {\bibinfo {author} {\bibfnamefont {H.-H.}\ \bibnamefont
  {Teh}}, \bibinfo {author} {\bibfnamefont {W.}~\bibnamefont {Dou}}, \ and\
  \bibinfo {author} {\bibfnamefont {J.~E.}\ \bibnamefont {Subotnik}},\
  }\bibfield  {title} {\enquote {\bibinfo {title} {Antisymmetric berry
  frictional force at equilibrium in the presence of spin-orbit coupling},}\
  }\href {\doibase 10.1103/PhysRevB.104.L201409} {\bibfield  {journal}
  {\bibinfo  {journal} {Phys. Rev. B}\ }\textbf {\bibinfo {volume} {104}},\
  \bibinfo {pages} {L201409} (\bibinfo {year} {2021})}\BibitemShut {NoStop}%
\bibitem [{\citenamefont {Teh}, \citenamefont {Dou},\ and\ \citenamefont
  {Subotnik}(2022)}]{PhysRevB.106.184302}%
  \BibitemOpen
  \bibfield  {author} {\bibinfo {author} {\bibfnamefont {H.-H.}\ \bibnamefont
  {Teh}}, \bibinfo {author} {\bibfnamefont {W.}~\bibnamefont {Dou}}, \ and\
  \bibinfo {author} {\bibfnamefont {J.~E.}\ \bibnamefont {Subotnik}},\
  }\bibfield  {title} {\enquote {\bibinfo {title} {Spin polarization through a
  molecular junction based on nuclear berry curvature effects},}\ }\href
  {\doibase 10.1103/PhysRevB.106.184302} {\bibfield  {journal} {\bibinfo
  {journal} {Phys. Rev. B}\ }\textbf {\bibinfo {volume} {106}},\ \bibinfo
  {pages} {184302} (\bibinfo {year} {2022})}\BibitemShut {NoStop}%
\bibitem [{\citenamefont {Wu}\ and\ \citenamefont
  {Subotnik}(2021)}]{NatComms.12.700}%
  \BibitemOpen
  \bibfield  {author} {\bibinfo {author} {\bibfnamefont {Y.}~\bibnamefont
  {Wu}}\ and\ \bibinfo {author} {\bibfnamefont {J.~E.}\ \bibnamefont
  {Subotnik}},\ }\bibfield  {title} {\enquote {\bibinfo {title} {Electronic
  spin separation induced by nuclear motion near conical intersections},}\
  }\href {\doibase 10.1038/s41467-020-20831-8} {\bibfield  {journal} {\bibinfo
  {journal} {Nature Communications}\ }\textbf {\bibinfo {volume} {12}},\
  \bibinfo {pages} {700} (\bibinfo {year} {2021})}\BibitemShut {NoStop}%
\bibitem [{\citenamefont {Tao}, \citenamefont {Qiu},\ and\ \citenamefont
  {Subotnik}(2023)}]{JPhysChemLett.14.770}%
  \BibitemOpen
  \bibfield  {author} {\bibinfo {author} {\bibfnamefont {Z.}~\bibnamefont
  {Tao}}, \bibinfo {author} {\bibfnamefont {T.}~\bibnamefont {Qiu}}, \ and\
  \bibinfo {author} {\bibfnamefont {J.~E.}\ \bibnamefont {Subotnik}},\
  }\bibfield  {title} {\enquote {\bibinfo {title} {Symmetric post-transition
  state bifurcation reactions with berry pseudomagnetic fields},}\ }\href
  {\doibase 10.1021/acs.jpclett.2c02668} {\bibfield  {journal} {\bibinfo
  {journal} {The Journal of Physical Chemistry Letters}\ }\textbf {\bibinfo
  {volume} {14}},\ \bibinfo {pages} {770--778} (\bibinfo {year}
  {2023})}\BibitemShut {NoStop}%
\bibitem [{\citenamefont {Chen}\ and\ \citenamefont
  {Subotnik}(2023)}]{JPhysChemLett.14.5665}%
  \BibitemOpen
  \bibfield  {author} {\bibinfo {author} {\bibfnamefont {J.}~\bibnamefont
  {Chen}}\ and\ \bibinfo {author} {\bibfnamefont {J.}~\bibnamefont
  {Subotnik}},\ }\bibfield  {title} {\enquote {\bibinfo {title} {Nonadiabatic
  potential energy surfaces for a molecule on a surface as found by constrained
  complete active space theory},}\ }\href {\doibase
  10.1021/acs.jpclett.3c00777} {\bibfield  {journal} {\bibinfo  {journal} {The
  Journal of Physical Chemistry Letters}\ }\textbf {\bibinfo {volume} {14}},\
  \bibinfo {pages} {5665--5673} (\bibinfo {year} {2023})}\BibitemShut {NoStop}%
\bibitem [{\citenamefont {Gupta}\ \emph
  {et~al.}(2023{\natexlab{b}})\citenamefont {Gupta}, \citenamefont {Kumar},
  \citenamefont {Bhowmick}, \citenamefont {Fontanesi}, \citenamefont {Paltiel},
  \citenamefont {Fransson},\ and\ \citenamefont
  {Naaman}}]{JPhysChemLett.14.9377}%
  \BibitemOpen
  \bibfield  {author} {\bibinfo {author} {\bibfnamefont {A.}~\bibnamefont
  {Gupta}}, \bibinfo {author} {\bibfnamefont {A.}~\bibnamefont {Kumar}},
  \bibinfo {author} {\bibfnamefont {D.~K.}\ \bibnamefont {Bhowmick}}, \bibinfo
  {author} {\bibfnamefont {C.}~\bibnamefont {Fontanesi}}, \bibinfo {author}
  {\bibfnamefont {Y.}~\bibnamefont {Paltiel}}, \bibinfo {author} {\bibfnamefont
  {J.}~\bibnamefont {Fransson}}, \ and\ \bibinfo {author} {\bibfnamefont
  {R.}~\bibnamefont {Naaman}},\ }\bibfield  {title} {\enquote {\bibinfo {title}
  {Does coherence affect the multielectron oxygen reduction reaction?}}\ }\href
  {\doibase 10.1021/acs.jpclett.3c02594} {\bibfield  {journal} {\bibinfo
  {journal} {The Journal of Physical Chemistry Letters}\ }\textbf {\bibinfo
  {volume} {14}},\ \bibinfo {pages} {9377--9384} (\bibinfo {year}
  {2023}{\natexlab{b}})}\BibitemShut {NoStop}%
\bibitem [{\citenamefont {Fransson}\ and\ \citenamefont
  {Naaman}(2025)}]{JPhysChemLett.16.1629}%
  \BibitemOpen
  \bibfield  {author} {\bibinfo {author} {\bibfnamefont {J.}~\bibnamefont
  {Fransson}}\ and\ \bibinfo {author} {\bibfnamefont {R.}~\bibnamefont
  {Naaman}},\ }\bibfield  {title} {\enquote {\bibinfo {title} {Chirality
  assisted triplet electron pairing},}\ }\href {\doibase
  10.1021/acs.jpclett.4c03734} {\bibfield  {journal} {\bibinfo  {journal} {The
  Journal of Physical Chemistry Letters}\ }\textbf {\bibinfo {volume} {16}},\
  \bibinfo {pages} {1629--1633} (\bibinfo {year} {2025})}\BibitemShut {NoStop}%
\bibitem [{\citenamefont {Briggeman}\ \emph {et~al.}(2025)\citenamefont
  {Briggeman}, \citenamefont {Mansfield}, \citenamefont {Kombe}, \citenamefont
  {Damanet}, \citenamefont {Lee}, \citenamefont {Tang}, \citenamefont {Yu},
  \citenamefont {Biswas}, \citenamefont {Li}, \citenamefont {Huang},
  \citenamefont {Eom}, \citenamefont {Irvin}, \citenamefont {Daley},\ and\
  \citenamefont {Levy}}]{SciAdv.11.eadx4761}%
  \BibitemOpen
  \bibfield  {author} {\bibinfo {author} {\bibfnamefont {M.}~\bibnamefont
  {Briggeman}}, \bibinfo {author} {\bibfnamefont {E.}~\bibnamefont
  {Mansfield}}, \bibinfo {author} {\bibfnamefont {J.}~\bibnamefont {Kombe}},
  \bibinfo {author} {\bibfnamefont {F.}~\bibnamefont {Damanet}}, \bibinfo
  {author} {\bibfnamefont {H.}~\bibnamefont {Lee}}, \bibinfo {author}
  {\bibfnamefont {Y.}~\bibnamefont {Tang}}, \bibinfo {author} {\bibfnamefont
  {M.}~\bibnamefont {Yu}}, \bibinfo {author} {\bibfnamefont {S.}~\bibnamefont
  {Biswas}}, \bibinfo {author} {\bibfnamefont {J.}~\bibnamefont {Li}}, \bibinfo
  {author} {\bibfnamefont {M.}~\bibnamefont {Huang}}, \bibinfo {author}
  {\bibfnamefont {C.-B.}\ \bibnamefont {Eom}}, \bibinfo {author} {\bibfnamefont
  {P.}~\bibnamefont {Irvin}}, \bibinfo {author} {\bibfnamefont {A.~J.}\
  \bibnamefont {Daley}}, \ and\ \bibinfo {author} {\bibfnamefont
  {J.}~\bibnamefont {Levy}},\ }\bibfield  {title} {\enquote {\bibinfo {title}
  {Engineered chirality of one-dimensional nanowires},}\ }\href {\doibase
  10.1126/sciadv.adx4761} {\bibfield  {journal} {\bibinfo  {journal} {Science
  Advances}\ }\textbf {\bibinfo {volume} {11}},\ \bibinfo {pages} {eadx4761}
  (\bibinfo {year} {2025})},\ \Eprint
  {http://arxiv.org/abs/https://www.science.org/doi/pdf/10.1126/sciadv.adx4761}
  {https://www.science.org/doi/pdf/10.1126/sciadv.adx4761} \BibitemShut
  {NoStop}%
\end{thebibliography}%

\end{document}